\documentclass[12pt]{iopart}

\usepackage{amsmath}
\usepackage{amssymb}
\usepackage{color}
\usepackage{adjustbox}
\usepackage{float}
\usepackage{verbatim}
\usepackage[bookmarks]{hyperref}
\usepackage{enumerate}
\usepackage{amsthm}
\usepackage{graphicx}
\usepackage[english]{babel}
\usepackage{cite}
\usepackage{braket}
\usepackage{subcaption}
\usepackage{booktabs}
\usepackage{tikz}
\usepackage{tikz-3dplot}
\usepackage[normalem]{ulem}

\begin{document}
\title[]{Bayesian approach to simultaneous quantum multiparameter estimation with finite data}
\author{Chun Kit Dennis Law}
\address{Forschungszentrum J\"ulich, Institute of Quantum Control (PGI-8), 52425 J\"ulich, Germany}
\author{J\'ozsef Zsolt Bern\'ad \footnote[7]{Author to whom any correspondence should be addressed.}}
\address{HUN-REN Wigner Research Centre for Physics, Konkoly-Thege M. \'{u}t 29-33, H-1121 Budapest, Hungary}
\ead{bernad.jozsef.zsolt@wigner.hu}

\date{today}

\begin{abstract}
Multiparameter estimation remains a fundamental challenge in quantum estimation theory because the optimal measurements associated with different parameters are generally incompatible. In this work, we develop a Bayesian framework for the simultaneous estimation of multiple parameters. Our approach builds on Personick's method, in which the estimation problem is reduced to a set of Lyapunov equations whose solutions define optimal observables for the individual parameters. Since these observables generally do not commute, we construct a convex combination of the solutions, parameterized by a set of variational parameters. The spectral decomposition of the resulting operator defines a parametrized projection-valued measure. The resulting projective measurement determines the likelihood function, from which the minimum mean-square error estimators and the corresponding Bayesian mean-square errors are obtained, following standard Bayesian procedures, as functions of the variational parameters. To determine their optimal values, we formulate a minimax optimization problem based on the normalized Bayesian mean-square errors, introducing a min–max normalization procedure that enables a meaningful comparison of estimation errors associated with different parameters. This optimization yields a variationally optimized projective measurement for simultaneous Bayesian estimation. Two qubit examples, involving phase estimation and the estimation of parameters defining convex combinations of unitary operations, demonstrate the construction of optimized projective measurements and the corresponding minimum mean-square error estimators within the proposed framework for finite data sets.
\end{abstract}
\vspace{2pc}

\maketitle

\section{Introduction}
Parameter estimation is a well-established branch of statistical inference with extensive applications in science and engineering \cite{VanTrees1968, vandenbosbbook, Kay2009}. Its objective is to estimate unknown parameters by extracting information from the measurement data and quantifying the associated uncertainty. In the context of quantum systems, information about the parameters of interest is obtained through quantum measurements.  Naturally, optimizing both the measurement process and the extraction of information from the resulting data is crucial for the efficient characterization of quantum systems and their dynamics. This area of study, known as quantum statistical inference, plays a fundamental role in many quantum information processing tasks. Originally developed in the mid-1960s \cite{helstrombook, holevobook}, it has since developed into a vibrant and active field of research with numerous practical applications \cite{ParisRehacek2004, hayashi_asymptotic, Hayashi2017, SidhuKok2020}. Although single parameter estimation of quantum systems has already found ample technological applications \cite{Marciniak2022, Rubio2021, Rubio2022, Pedrozo2020, Robinson2024, Ganapathy2023}, models consisting of only a single unknown parameter are often over-simplifications of real quantum systems of technological interest \cite{Szczykulska2016}. Problems involving simultaneous estimation of multiple parameters are ubiquitous in quantum technology. Examples include phase and dephasing coefficients in atomic interferometry \cite{Crowley2014, Gessner2018}, waveform estimation \cite{Tsang2011}, quantum imaging \cite{Tsang2016, Lupo2016, Chrostowski2017, Zhou2019}, vector-field sensors \cite{Baumgratz2016, Kaubruegger2023}, and optomechanical systems \cite{Sanavio2022}. 

A general quantum estimation problem can be naturally decomposed into two stages: (i) the identification of an optimal measurement strategy, and (ii) the construction of classical estimators based on the resulting measurement data \cite{Hayashi2017}. Despite extensive developments in quantum estimation theory, much of the existing literature has concentrated on characterizing the precision limits of estimator performance within the frequentist framework \cite{Paris2009, Wang2024, Holdsworth2025}. However, comprehensive overviews of precision bounds in both frequentist and Bayesian paradigms are available in \cite{Demkowicz2020, Li2018}. Among the various bounds, the Cramér–Rao bound admits a relatively direct extension to quantum systems via the symmetric logarithmic derivative \cite{Helstrom1968}. More general formulations include quantum Cramér–Rao bounds based on right and left logarithmic derivatives \cite{Yuen, Belavkin}, as well as the Holevo Cramér–Rao bound \cite{Holevo}. The latter is of central importance in multiparameter estimation, which can often involve incompatible optimal measurements for different parameters \cite{Ragy, Carollo2019}. While the Holevo Cramér–Rao bound is saturable for pure states \cite{Matsumoto}, in the case of mixed states it can generally only be attained asymptotically \cite{HayashiMat, Kahn, Yamagata}. A key limitation of these bounds is their reliance on local unbiasedness and the assumption of asymptotically large data sets. In realistic experimental scenarios, however, acquiring large amounts of data may be impractical, for instance due to constraints on optical power and photon number in quantum optical implementations \cite{Berchera2019, Polino2019}. In such cases with limited data, Bayesian approaches provide an alternative by incorporating prior information. It is worth noting that small sample frequentist methodologies also exist in classical statistics \cite{Demidenko}. In the quantum domain, optimal Bayesian strategies have been studied since the seminal works of Helstrom, Holevo, and Personick \cite{helstrombook, holevobook, Personick1970, Personick1971}, and continue to attract significant attention \cite{Machiavello2003, Rubio2020, Kaubruegger2023, Lee2022}. These approaches focus on determining measurement schemes that minimize a chosen Bayesian risk, which can be achieved analytically in certain cases, such as phase estimation with pure states \cite{Machiavello2003}. More recently, efforts have also been directed toward deriving practical Bayesian bounds and identifying corresponding optimal measurements. Nevertheless, the fundamental issue of measurement incompatibility in multiparameter estimation remains a significant challenge in this framework \cite{Rubio2020}.
 
The aim of this work is to develop an optimal estimation strategy for multiparameter quantum systems. Our approach is based on the construction of optimal classical estimators as functions of the measurement data, an aspect that has received comparatively limited attention in the quantum metrology literature. As shown in the single parameter scenarios presented in \cite{Law2025}, parametric models \cite{Wasserman} arising in quantum mechanics may fail to be identifiable \cite{Casella}. Consequently, they can exhibit ambiguities in commonly used estimators, including those that are asymptotically efficient. A notable exception is the minimum mean-square error (MMSE) estimator within the Bayesian framework. In classical statistical inference, the MMSE estimator is known to be optimal not only for finite data but also in multiparameter estimation settings, as it minimizes the Bayesian mean-square error (BMSE) \cite{Kay2009, VanTrees1968}. These advantages motivate us to address multiparameter estimation problems in quantum systems using MMSE estimators. While this possibility has been recognized in the literature, existing approaches exhibit certain limitations. In particular, some works do not analyze the individual performance of the different estimators, as the optimization is performed with respect to the sum of the BMSEs of the estimators \cite{Kaubruegger2023}. Other approaches reduce the multiparameter estimation problem to an effective single-parameter problem by considering linear combinations of the parameters of interest, thereby circumventing the issue of measurement incompatibility \cite{Lee2022}. The approach of Ref. \cite{Kaubruegger2023} is closer in spirit to our objective; however, our aim is to develop a strategy that explicitly incorporates the individual performance of each estimator into the optimization procedure. This distinction is important because, for a given parametric model, dataset, and prior probability density function, the estimators produce concrete numerical values, i.e., estimates, whose magnitudes may differ significantly. Consequently, optimizing the sum of the BMSEs tends to prioritize the precision of the estimates contributing most to the total BMSE, potentially at the expense of those with smaller, yet still non-negligible, contributions.

Our approach is based on the framework developed by Personick, in which the estimation problem is reduced to a set of Lyapunov equations associated with the parameters of interest \cite{Personick1970, Personick1971}. The solutions of these equations are optimal self-adjoint operators whose eigenprojectors define the corresponding optimal measurements, while their eigenvalues provide the corresponding single-shot estimates. In general, however, these operators do not commute and therefore do not share a common eigenbasis, precluding their simultaneous measurement. To overcome this limitation, we construct a convex combination of the operator solutions by introducing a set of variational parameters. The eigendecomposition of the resulting operator defines a parameterized projection-valued measure, which in turn induces a likelihood function. Combined with a chosen prior distribution, this likelihood yields a posterior distribution and a joint distribution of the data and the parameters to be estimated. Standard Bayesian procedures \cite{Kay2009} are then used to derive the MMSE estimators and the corresponding BMSEs as functions of the introduced variational parameters. Finally, the variational parameters are selected according to a minimax framework \cite{Ethier2010}, namely by minimizing the maximum of the BMSEs. Since the BMSEs associated with different parameters may differ substantially in magnitude, we introduce a min–max normalization procedure \cite{Han2012, Aggarwal2015}, a standard
technique originating from scaling in data analysis and data mining, to place them on a common scale before optimization. This optimization determines the variational parameters and, consequently, the optimal projective measurement. The resulting framework provides a general strategy for Bayesian multiparameter estimation of mixed quantum states, yielding a single optimized projective measurement scheme that enables the simultaneous inference of all unknown parameters from a finite data set of measurement outcomes.

The article is organized as follows. In Sec. \ref{sec:preliminaries}, we introduce the basic theoretical framework of quantum estimation theory within the Bayesian paradigm. We then outline the key steps in the derivation of optimal estimators for the parameters of interest, following the approach developed in \cite{Personick1971, Demkowicz2020}. In Sec. \ref{sec:nouveltechnique}, we present our method for the estimation of equally weighted parameters based on convex combinations and post-processing of classical data. The proposed strategy is illustrated in Sec. \ref{sec:applications} through two examples that involve two-level systems with two unknown parameters. Finally, in Sec. \ref{sec:conluding}, we summarize our findings and present our conclusions.

\section{Preliminaries}
\label{sec:preliminaries}

In this section, before presenting our approach and results, we briefly introduce the quantum estimation framework employed in this work. This provides the necessary background and fixes the notation used throughout the paper.

{\it Bayesian risk and the minimum mean-square error estimator}. Let $\mathcal{H}$ be a $d$-dimensional Hilbert space and let $\Theta \subset \mathbb{R}^p$ denote the parameter space of the $p$ unknown parameters. For a vector of unknown parameters $\boldsymbol{\theta} = (\theta_{1}, \theta_{2}, \cdots, \theta_{p})^{T}$ ($T$ denotes the transposition), a quantum parametric model is defined as a family of density operators $\{\rho_{\boldsymbol{\theta}} \vert \boldsymbol{\theta} \in \Theta \}$ defined on $\mathcal{H}$. In this framework, the mapping $\boldsymbol{\theta} \mapsto \rho_{\boldsymbol{\theta}}$ is implicitly assumed to be smooth, thereby ensuring that derivatives of the density operator with respect to the parameters are well defined. A general measurement on a quantum system is described by a positive operator-valued measure (POVM), that is, a collection of positive semidefinite operators $\{\Pi_{x}\}_{x \in \mathcal{X}}$, i.e., $\Pi_{x} \geq 0$ for all $x \in \mathcal{X}$, satisfying $\sum_{x \in \mathcal{X}} \Pi_{x} = I_{d}$ with $I_{d}$ being the $d$-dimensional identity matrix on the Hilbert space $\mathcal{H}$. For two self-adjoint operators $A$ and $B$ on $\mathcal{H}$, the notation $B \geq A$ means that $B-A \geq 0$, i.e., $B-A$ is a positive semidefinite operator. 

Measurement outcomes are inherently random and induce a classical random variable $X:\Omega \mapsto \mathcal{X}$, where $\Omega$
denotes the sample space and $\mathcal{X} \subseteq \mathbb{R}$ is the set of possible measurement outcomes. We denote the conditional probability mass function (PMF) for measuring the outcome $x$ given the unknown parameters $\boldsymbol{\theta}$ as $p(x \vert \boldsymbol{\theta})=\Tr[\rho_{\boldsymbol{\theta}}\Pi_{x}]$. Here, $\Tr[\cdot]$ denotes the trace over the Hilbert space $\mathcal{H}$. In the following, this will be referred to as the likelihood function. A special class of measurements is given by projection-valued measures (PVMs), which are POVMs whose elements are orthogonal projections. In this work, we restrict the proposed method to PVMs.

In the Bayesian framework, the unknown parameter vector $\boldsymbol{\theta}$ is itself treated as a random variable, with prior information encoded in a prior probability density function (PDF) $\pi(\boldsymbol{\theta})$. Among the various Bayesian estimation strategies, we restrict our attention to point estimation. Accordingly, we consider estimators of the form
\begin{equation}
    \hat{\boldsymbol{\theta}}(\mathbf{x}) = \left(\hat{\theta}_{1}(\mathbf{x}), \hat{\theta}_{2}(\mathbf{x}), \cdots, \hat{\theta}_{p}(\mathbf{x})\right)^{T},
\end{equation}
which depend on the experimentally obtained data set $\mathbf{x}=\{x_1, x_2, \cdots, x_N\}$ consisting of $N$ measurement outcomes. Optimal estimators are defined as those that minimize the Bayesian risk
\begin{equation}\label{eqn:BayesRiskDef}
    \mathcal{R}^{(N)} = \sum_{\mathbf{x}} \int_\Theta \mathcal{C}\big[\hat{\boldsymbol{\theta}}(\mathbf{x}), \boldsymbol{\theta}\big] p(\mathbf{x}, \boldsymbol{\theta})\,d\boldsymbol{\theta},
\end{equation}
where $\mathcal{C}[\cdot,\cdot]$ is a specified cost function and the joint PDF is given by 
\begin{equation}
 p(\mathbf{x}, \boldsymbol{\theta}) = p(\mathbf{x} \vert \boldsymbol{\theta}) \pi(\boldsymbol{\theta}).   
\end{equation}
The sum over $\mathbf{x}$ denotes summation over all possible values of $x_1, x_2, \dots x_N$ from the set $\mathcal{X}$.

In practice, experimentalists perform repeated measurements to reduce the statistical uncertainty in the estimates of quantities of interest. For classical systems, measurements typically do not disturb the system, allowing repeated observations without the need for reinitialization. In contrast, quantum systems are highly sensitive to measurements: each measurement generally perturbs the state of the system. As a result, successive measurements on the same quantum system produce outcomes that are conditioned on previous measurements. For this reason, meaningful repeated measurements on a quantum system necessitates the reinitialization procedure. A common simplifying assumption is that the observed data due to repeated measurements consist of independent and identically distributed (i.i.d.) random variables. We also adopt this assumption in this work, although it does not hold in general \cite{Casella, Hampel, Li}. Under this assumption, each experimental run is regarded as an independent copy of the system, prepared using the same state-preparation procedure. The resulting $N$ copies are therefore uncorrelated, and the joint state can be written as
\begin{equation}
\label{eq:Ntensor}
    \rho_{\boldsymbol{\theta}}^{\otimes N} = \underbrace{\rho_{\boldsymbol{\theta}} \otimes \rho_{\boldsymbol{\theta}} \otimes \cdots \otimes \rho_{\boldsymbol{\theta}}}_{N\text{-fold tensor product}}.
\end{equation}
In this setting, a PVM is defined on the composite Hilbert space $\mathcal{H}^{\otimes N}$ and is assumed to factorize in tensor-product form, leading to a factorized likelihood function
\begin{equation}
\label{eq: likelihoodf}
    p(\mathbf{x} \vert \boldsymbol{\theta})=\prod_{i=1}^N p(x_i \vert \boldsymbol{\theta}).
\end{equation}
For the cost function
\begin{equation}\label{eqn:BayesRiskApprox}
    \mathcal{C}\big[\hat{\boldsymbol{\theta}}(\mathbf{x}), \boldsymbol{\theta}\big] =\left( \hat{\boldsymbol{\theta}}(\mathbf{x})-\boldsymbol{\theta} \right)^{T}\cdot \left( \hat{\boldsymbol{\theta}}(\mathbf{x}) -\boldsymbol{\theta} \right),
\end{equation}
which corresponds to the sum of squared errors, the Bayesian risk defined in \eqref{eqn:BayesRiskDef} reduces to the Bayesian mean-square error (BMSE), which is 
minimized by the vector minimum mean-square error (MMSE) estimator \cite{VanTrees1968} 
\begin{equation}
\label{eq:MMSEcl}
    \hat{\boldsymbol{\theta}}_{\text{\tiny{MMSE}}}(\mathbf{x}) = \int_\Theta \boldsymbol{\theta}\, p(\boldsymbol{\theta} \vert \mathbf{x}) \, d\boldsymbol{\theta},
\end{equation}
where 
\begin{equation}
\label{eq:posteriorcl}
    p(\boldsymbol{\theta} \vert \mathbf{x}) = \frac{p(\mathbf{x} \vert \boldsymbol{\theta}) \pi(\boldsymbol{\theta})}{\int_{\Theta} p(\mathbf{x} \vert \boldsymbol{\theta}) \pi(\boldsymbol{\theta}) \, d \boldsymbol{\theta}} 
\end{equation}
is the posterior probability. Furthermore, we emphasize that Eq.~\eqref{eqn:BayesRiskApprox} is only well-defined under the condition that all parameters to be estimated have the same physical dimensionality, since it is not meaningful to sum quantities with different units. This is readily understood in physics, where, for example, one might simultaneously estimate the temperature and the frequency of a system. If the parameters have different dimensions, suitable dimensionless variables must be introduced in order to properly define the BMSE.

Although this generalization is not explicitly discussed in Refs. \cite{VanTrees1968} and \cite{Kay2009}, the same proof can be used to derive \eqref{eq:MMSEcl} for the general quadratic cost function
\begin{equation}
\label{eqn:costmatrix}
\mathcal{C}\big[\hat{\boldsymbol{\theta}}
(\mathbf{x}), \boldsymbol{\theta}\big] =\left(\hat{\boldsymbol{\theta}}(\mathbf{x})-\boldsymbol{\theta} \right)^{T} C \left( \hat{\boldsymbol{\theta}}(\mathbf{x})-\boldsymbol{\theta} \right),    
\end{equation}
where $C$ is the cost matrix, i.e., a real positive definite matrix. Therefore, the cost matrix $C$ does not affect the optimal classical MMSE estimators. In the subsequent section, we formulate Personick’s variational approach, originally developed for single-parameter quantum estimation \cite{Personick1971}, directly in the multiparameter setting \cite{Demkowicz2020}. We show that $C$ likewise does not enter the optimization of the quantum measurements. Thus, within this variational framework, the optimization of the measurement is independent of the relative weighting of the parameter estimation errors encoded in $C$.

{\it Personick's approach and the fundamental problem in multiparameter estimation.}
In this part, we recall the variational approach pioneered by Personick \cite{Personick1970, Personick1971} and present the result necessary for our approach, based on Ref.~\cite{Demkowicz2020}. Our starting point is a single copy of a density operator $\rho_{\boldsymbol{\theta}}$. The extension to the $N$-copy setting, together with a detailed derivation, is presented in Appendix~\ref{sec:AppendixA}.
The BMSE for a classical estimation problem based on a single measurement outcome is given by
\begin{equation}
    \mathcal{R}^{(1)} = \sum_{x \in \mathcal{X}} \int_\Theta \left(\hat{\boldsymbol{\theta}}(x)-\boldsymbol{\theta} \right)^{T} C \left(\hat{\boldsymbol{\theta}}(x)-\boldsymbol{\theta} \right)p(x \vert \boldsymbol{\theta}) \pi(\boldsymbol{\theta})\, d\boldsymbol{\theta}.
\end{equation}
In the quantum setting, as introduced previously, $p(x \vert \boldsymbol{\theta}) =\Tr[\rho_{\boldsymbol{\theta}}\Pi_{x}]$. This allows us to express the BMSE as
\begin{equation}
    \mathcal{R}^{(1)} = \sum_{x \in \mathcal{X}} \int_\Theta \left(\hat{\boldsymbol{\theta}}(x)-\boldsymbol{\theta} \right)^{T} C \left(\hat{\boldsymbol{\theta}}(x)-\boldsymbol{\theta} \right) \Tr \left[ \rho_{\boldsymbol{\theta}} \Pi_{x} \right] \pi(\boldsymbol{\theta})\, d\boldsymbol{\theta}.
\end{equation}
Upon the introduction of the notations 
\begin{eqnarray}
\label{eq: Gammas}
    \boldsymbol{\Lambda}_{1}&=&\sum_{x \in \mathcal{X}} \hat{\boldsymbol{\theta}}(x) \Pi_{x}, \quad \Lambda_{2}=\sum_{x \in \mathcal{X}} \hat{\boldsymbol{\theta}}^T(x) C \hat{\boldsymbol{\theta}}(x) \Pi_{x},  \\
    \Gamma_{0}&=&\int_\Theta \rho_{\boldsymbol{\theta}} \pi (\boldsymbol{\theta})\, d\boldsymbol{\theta}, \quad \boldsymbol{\Gamma}_{1}=\int_\Theta \boldsymbol{\theta} \rho_{\boldsymbol{\theta}} \pi (\boldsymbol{\theta})\, d\boldsymbol{\theta}, \quad \Delta_{C}^{2} \boldsymbol{\theta} = \int_\Theta \boldsymbol{\theta}^{T} C \boldsymbol{\theta} \pi (\boldsymbol{\theta})\, d\boldsymbol{\theta}, \nonumber
\end{eqnarray}
where $\boldsymbol{\Lambda}_{1}$ and $\boldsymbol{\Gamma}_{1}$ are $p$-dimensional vectors with self-adjoint operator-valued entries, the BMSE reads
\begin{equation}\label{Cost}
    \mathcal{R}^{(1)} = \Delta_{C}^{2} \boldsymbol{\theta} - 2 \Tr \left[ \boldsymbol{\Gamma}_{1}^{T} C \boldsymbol{\Lambda}_{1} \right] + \Tr[\Gamma_{0}\Lambda_{2}].
\end{equation}
Furthermore, $\Gamma_0$ is a self-adjoint operator and $\Lambda_2$ a positive semidefinite operator. One can observe that
\begin{equation}
 \sum_{x \in \mathcal{X}} \left(\hat{\boldsymbol{\theta}}(x)-\boldsymbol{\Lambda}_{1} \right)^{T} C \left(\hat{\boldsymbol{\theta}}(x) -\boldsymbol{\Lambda}_{1} \right) \Pi_{x}  \geq 0,  
\end{equation}
which follows directly from the positive semidefiniteness of  $\Pi_{x}$ and the fact that the cost matrix $C$ is real and positive definite. This implies that 
\begin{equation}
   \Lambda_{2} \geq \boldsymbol{\Lambda}_{1}^{T} C \boldsymbol{\Lambda}_{1}. 
\end{equation}
Using this result, the BMSE is bounded from below by
\begin{eqnarray}
    \mathcal{R}^{(1)} \geq B(\boldsymbol{\Lambda}_{1}) &:= \Delta_{C}^{2} \boldsymbol{\theta} - 2 \Tr \left[ \boldsymbol{\Gamma}_{1}^{T} C \boldsymbol{\Lambda}_{1} \right] + \Tr[\Gamma_{0}\boldsymbol{\Lambda}_{1}^{T} C \boldsymbol{\Lambda}_{1}] \notag \\
    &= \Delta_{C}^{2} \boldsymbol{\theta} + \tr\left[ C \left\{ \Tr[\Gamma_{0}\boldsymbol{\Lambda}_{1} \boldsymbol{\Lambda}_{1}^{T}] - 2 \Tr \left[ \boldsymbol{\Lambda}_{1} \boldsymbol{\Gamma}_{1}^{T} \right] \right\} \right] ,
\end{eqnarray}
where we have introduced the trace operation $\tr[\cdot]$, acting on the $p$-dimensional parameter space $\Theta$. The expressions $\boldsymbol{\Lambda}_{1} \boldsymbol{\Lambda}_{1}^{T}$ and $\boldsymbol{\Lambda}_{1} \boldsymbol{\Gamma}_{1}^{T}$ denote $p \times p$ matrices with operator-valued entries that act as linear operators on the Hilbert space $\mathcal{H}$. 

Now, let $\epsilon$ be a real number and $\boldsymbol{H}$ any $p$-dimensional vector with self-adjoint operator-valued entries. Let $\boldsymbol{\Lambda}_{\text{opt}}$ be the $p$-dimensional vector with self-adjoint operator-valued entries that minimizes $B(\boldsymbol{\Lambda}_{1})$. Then we have
\begin{equation}
 B(\boldsymbol{\Lambda}_{\text{opt}}) \leqslant B (\boldsymbol{\Lambda}_{\text{opt}}+ \epsilon \boldsymbol{H}).
\end{equation}
Evaluating the right-hand side of this inequality and differentiating with respect to $\epsilon$ at $\epsilon=0$ yields a Lyapunov equation in vector form:
\begin{equation}
    \boldsymbol{\Lambda}_{\text{opt}} \Gamma_{0} + \Gamma_{0} \boldsymbol{\Lambda}_{\text{opt}} = 2\boldsymbol{\Gamma}_{1},
\end{equation}
whose solution, provided that $\Gamma_0>0$, is given by \cite{Personick1971}
\begin{equation}\label{Solution1}
    \boldsymbol{\Lambda}_{\text{opt}} = 2 \int_{0}^{\infty} e^{-\Gamma_{0} y} \boldsymbol{\Gamma}_{1} e^{-\Gamma_{0} y} dy.
\end{equation} 
The above minimization of $B(\boldsymbol{\Lambda}_{1})$ allows one to bound $\mathcal{R}^{(1)}$ from below as
\begin{equation}\label{eqn:finalbound}
    \mathcal{R}^{(1)} \geq \Delta_{C}^{2} \boldsymbol{\theta} - \Tr \left[\boldsymbol{\Lambda}_{\text{opt}}^{T} \Gamma_{0} \boldsymbol{\Lambda}_{\text{opt}} \right] .
\end{equation} 
The first term on the right-hand side of Eq. \eqref{eqn:finalbound} represents the uncertainty associated with the prior, while the second term accounts for the error arising from the quantum estimation procedure.

The derivation of $\boldsymbol{\Lambda}_{\text{opt}}$ reveals a fundamental problem in multi-parameter estimation of quantum systems. It is important to recall that the solution to the Lyapunov equation in Eq. \eqref{Solution1} is a $p$-dimensional vector in the parameter space $\Theta$, with its $i$-th entry $\Lambda_{\text{opt},i}$ being a self-adjoint operator acting on the Hilbert space $\mathcal{H}$. The different entries of $\boldsymbol{\Lambda}_{\text{opt}}$ are not guaranteed to commute with each other, due to the possible non-commutativity between the entries of $\boldsymbol{\Gamma}_{1}$. This poses a serious challenge in quantum estimation, since there may not exist a common eigenbasis for the simultaneous measurements of all entries of $\boldsymbol{\Lambda}_{\text{opt}}$. If this scenario arises, it is inevitable that one must sacrifice the precision of the estimate of one variable to improve that of others, and in the subsequent discussion we present a possible resolution to this issue.

\section{Method based on convex combinations of Personick-type solutions}\label{sec:nouveltechnique}

In this section, we present a new approach for estimating $p$ parameters in the case where the entries of $\boldsymbol{\Lambda}_{\text{opt}}$ do not commute with one another, as discussed in Sec. \ref{sec:preliminaries}. The approach combines the optimization of convex combinations of Personick-type solutions with classical data post-processing, enabling the determination of a suitable measurement operator and the subsequent construction of the corresponding MMSE estimators for the unknown parameters. Although the preceding discussion applies to an arbitrary cost matrix $C$, this aspect is beyond the scope of the present work. Therefore, we restrict our method to the identity cost matrix, i.e., $C=I_{p}$.

\subsection{Construction of single projection-valued measure}

To construct a single measurement for the simultaneous estimation of all parameters, we combine the individual Personick-type solutions into a single operator. Since the entries of $\boldsymbol{\Lambda}_{\text{opt}}$ need not commute, there may not exist a projective measurement that is simultaneously optimal for estimating $\theta_i$ and $\theta_j$. We therefore consider convex combinations of the entries $\Lambda_{\text{opt},i}$, for $1 \leq i \leq p$. The resulting convex hull provides additional degrees of freedom for constructing a single PVM. By exploring this convex set, we seek a measurement that achieves a more balanced BMSE across the unknown parameters, since each $\Lambda_{\text{opt},i}$ is obtained by optimizing the estimation of the corresponding parameter $\theta_i$. To this end, we introduce a set of variational parameters
$\{\alpha_{i}\}_{i=1}^{p}$, where $\alpha_{i} \in [0,1]$ and $\sum_{i=1}^{p} \alpha_{i}=1$. The resulting convex combination of the entries of $\boldsymbol{\Lambda}_{\text{opt}}$ is given by
\begin{equation}\label{eqn:convex}
    M_{\boldsymbol{\alpha}} = \sum_{i=1}^{p} \alpha_{i} \Lambda_{ \text{opt},i}.
\end{equation} 
Here, $\boldsymbol{\alpha} = (\alpha_{1}, \alpha_{2}, \cdots , \alpha_{p})^{T}$ denotes the vector of variational parameters. Since each $\Lambda_{ \text{opt},i}$ is Hermitian and the coefficients $\alpha_i$ are real, $M_{\boldsymbol{\alpha}}$ is also Hermitian:
\begin{equation}
   M^\dagger_{\boldsymbol{\alpha}}=\left(\sum_{i=1}^{p} \alpha_{i} \Lambda_{ \text{opt},i}\right)^\dagger =\sum_{i=1}^{p} \alpha_{i} \Lambda^\dagger_{ \text{opt},i}=\sum_{i=1}^{p} \alpha_{i} \Lambda_{ \text{opt},i}=M_{\boldsymbol{\alpha}}.
\end{equation}
Consequently, $M_{\boldsymbol{\alpha}}$ admits the spectral decomposition
\begin{equation}
    M_{\boldsymbol{\alpha}}=\sum^d_{k=1} m_k P_{\boldsymbol{\alpha},k},
\end{equation}
where where $m_k$ denotes the $k$th eigenvalue of $M_{\boldsymbol{\alpha}}$, and $\{P_{\boldsymbol{\alpha},k}\}^d_{k=1}$ is the corresponding $\boldsymbol{\alpha}$-dependent PVM. This PVM is applied to the quantum system to generate the measurement outcomes used in the subsequent classical post-processing stage. The variational parameters $\boldsymbol{\alpha}$ are then determined by minimizing the BMSE of the resulting MMSE estimators, as discussed in the following sections.

\subsection{Classical data post-processing}\label{sec:dataprocessing}

After constructing the $\boldsymbol{\alpha}$-dependent PVM, the conditional PMF of obtaining the measurement outcome $k$ is given by
\begin{equation}
p_{\boldsymbol{\alpha}}(x=k \vert \boldsymbol{\theta}) = \Tr[\rho_{\boldsymbol{\theta}}P_{\boldsymbol{\alpha},k}],  
\end{equation}
as introduced in Sec.~\ref{sec:preliminaries}. Then, the likelihood function in Eq.~\eqref{eq: likelihoodf} also depends on $\boldsymbol{\alpha}$:
\begin{equation}
    p_{\boldsymbol{\alpha}}(\mathbf{x} \vert \boldsymbol{\theta})=\prod_{i=1}^N p_{\boldsymbol{\alpha}}(x_i \vert \boldsymbol{\theta})
\end{equation}
and the corresponding posterior distribution is given by
\begin{equation}
    p_{\boldsymbol{\alpha}}(\boldsymbol{\theta} \vert \mathbf{x}) = \frac{p_{\boldsymbol{\alpha}}(\mathbf{x} \vert \boldsymbol{\theta}) \pi(\boldsymbol{\theta})}{\int_{\Theta} p_{\boldsymbol{\alpha}}(\mathbf{x} \vert \boldsymbol{\theta}) \pi(\boldsymbol{\theta}) \, d \boldsymbol{\theta}}. 
\end{equation}
The $\boldsymbol{\alpha}$-dependent MMSE estimators read:
\begin{equation}                        \hat{\theta}_{\text{\tiny{MMSE}},i}({\bf x}, \boldsymbol{\alpha}) = \int_\Theta \theta_{i}\, p_{\boldsymbol{\alpha}}(\boldsymbol{\theta} \vert {\bf x}) \, d\boldsymbol{\theta},
\end{equation}
where $i\in\{1,2, \cdots p\}$. The BMSE associated with estimating $\theta_{i}$ from $N$ measurement outcomes takes the form
\begin{eqnarray}\label{eqn:TotalBMSE}
    \mathcal{R}^{(N)}_{i}(\boldsymbol{\alpha}) &=  \sum_{\mathbf{x}} \int d \boldsymbol{\theta} \left[\hat{\theta}_{\text{MMSE},i}(\mathbf{x}, \boldsymbol{\alpha}) - \theta_{i}\right]^{2} p_{\boldsymbol{\alpha}}(\mathbf{x}, \boldsymbol{\theta}) \nonumber \\
    &= \sum_{\mathbf{x}} p_{\boldsymbol{\alpha}}(\mathbf{x}) \left[ \int d \boldsymbol{\theta} \, \theta_{i}^{2} p_{\boldsymbol{\alpha}}(\boldsymbol{\theta} \vert \mathbf{x}) - \hat{\theta}^2_{\text{MMSE},i}(\mathbf{x}, \boldsymbol{\alpha})\right],
\end{eqnarray}
where we have used Bayes' theorem
\begin{equation}
 p_{\boldsymbol{\alpha}}(\mathbf{x}, \boldsymbol{\theta})=
 p_{\boldsymbol{\alpha}}(\mathbf{x}) p_{\boldsymbol{\alpha}}(\boldsymbol{\theta} \vert \mathbf{x})
\end{equation}
with
\begin{equation}
  p_{\boldsymbol{\alpha}}(\mathbf{x})= \int_{\Theta} p_{\boldsymbol{\alpha}}(\mathbf{x} \vert \boldsymbol{\theta}) \pi(\boldsymbol{\theta}) \, d \boldsymbol{\theta}.
\end{equation}
Since the choice $\alpha_{i}=1$ and $\alpha_{j}=0$ for all $j \neq i$ corresponds to using the Personick solution optimized specifically for estimating $\theta_{i}$, it follows that $\mathcal{R}^{(N)}_{i}(\boldsymbol{\alpha})$ attains its minimum at this point. An analogous argument applies to each of the remaining parameters. 

The next step is to devise a strategy for selecting suitable values of $\{\alpha_{i}\}_{i=1}^{p}$. A natural first approach is to determine the values of $\{\alpha_{i}\}_{i=1}^{p}$ that minimize the total BMSE $\mathcal{R}^{(N)}(\boldsymbol{\alpha})$. Since $C=I_{p}$, the total BMSE is given by
\begin{equation}
  \mathcal{R}^{(N)}(\boldsymbol{\alpha})=\sum^p_{i=1} \mathcal{R}^{(N)}_{i}(\boldsymbol{\alpha}).   
\end{equation}
However, this approach suffers from an inherent drawback. Since the ranges and magnitudes of the parameters $\{\theta_{i}\}_{i=1}^{p}$ may differ by several orders of magnitude, the total BMSE can be dominated by the contribution of one or a few parameters. Consequently, minimizing the total BMSE may effectively reduce to minimizing only the BMSE associated with the parameter having the largest error, thereby neglecting the estimation performance of the remaining parameters. An alternative criterion is therefore required. 

In general, the BMSEs associated with different parameters cannot be directly compared because their ranges and orders of magnitude may differ significantly. To overcome this issue, we employ a scaling technique known as min-max normalization \cite{Han2012, Aggarwal2015} and define the normalized BMSE
\begin{equation}
\label{eq:normalization}
    \eta_{i}^{(N)}(\boldsymbol{\alpha}) = \frac{\mathcal{R}^{(N)}_{i}(\boldsymbol{\alpha}) - \underset{\boldsymbol{\alpha}}{\min}\left[\mathcal{R}^{(N)}_{i}(\boldsymbol{\alpha})\right]}{\underset{\boldsymbol{\alpha}}{\max}\left[\mathcal{R}^{(N)}_{i}(\boldsymbol{\alpha})\right]-\underset{\boldsymbol{\alpha}}{\min}\left[\mathcal{R}^{(N)}_{i}(\boldsymbol{\alpha})\right]} ,
\end{equation}
where $i\in \{1, 2, \cdots, p\}$. By construction, the normalized BMSE satisfies $\eta^{(N)}_{i}(\boldsymbol{\alpha})  \in [0,1]$ for every parameter, regardless of its original range or magnitude. This normalization enables a meaningful comparison of the estimation performance across all parameters.

Note that $\eta^{(N)}_{i}(\boldsymbol{\alpha})$, like $\mathcal{R}^{(N)}_{i}(\boldsymbol{\alpha})$, attains its minimum at $\alpha_{i}=1$ and $\alpha_{j}=0$ for $j \neq i$. Since the min-max normalization places the normalized BMSEs of all parameters on an equal footing, no parameter should be systematically favored over the others during the optimization. We therefore adopt a minimax criterion by considering the surface
\begin{equation}
\tilde{\eta}^{(N)}(\boldsymbol{\alpha})=\max\left[\eta^{(N)}_{1}(\boldsymbol{\alpha}), \eta^{(N)}_{2}(\boldsymbol{\alpha}), \cdots , \eta^{(N)}_{p}(\boldsymbol{\alpha}) \right]    
\end{equation}
and seek the value of $\boldsymbol{\alpha}$ corresponding to its global minimum. We therefore adopt a minimax criterion by minimizing the maximum of the normalized BMSEs among all variational parameters \cite{Ethier2010}. In this way, the optimization seeks the best possible worst-case estimation performance, thereby preventing any single parameter from exhibiting an excessively large estimation error relative to the others. Equivalently, the optimized variational parameters are determined by
\begin{equation}
    \boldsymbol{\alpha}_{\text{opt}} = \underset{\boldsymbol{\alpha}}{\text{argmin}} \max \left[\eta^{(N)}_{1}(\boldsymbol{\alpha}), \eta^{(N)}_{2}(\boldsymbol{\alpha}), \cdots , \eta^{(N)}_{p}(\boldsymbol{\alpha}) \right].
\end{equation}
The resulting $\boldsymbol{\alpha}_{\text{opt}}$ is then substituted into the expressions for the MMSE estimators. Together with the measurement outcomes $\mathbf{x}$, this yields the optimized MMSE estimates $\hat{\theta}_{\text{\tiny{MMSE}},i}({\bf x}, \boldsymbol{\alpha}_{\text{opt}})$ and the corresponding quantum measurement described by the PVM $\{P_{\boldsymbol{\alpha}_{\text{opt}},k}\}^d_{k=1}$. The corresponding BMSEs, $\left\{R^{(N)}_{i}(\boldsymbol{\alpha}_{\text{opt}})\right\}_{i=1}^{p}$ are obtained accordingly.

While the extension to a general cost matrix is left for future investigation and lies beyond the scope of this article, the proposed methodology can be readily generalized to the case of a diagonal cost matrix with unequal weights. Suppose that the cost matrix is given by
\begin{equation}
 C = \text{diag}(w_1 ,w_{2},\cdots, w_{p}),   
\end{equation}
where $w_{i}>0$, subject to the constraint $\sum_{i=1}^{p} w_{i} =1$, denotes the weight assigned to the estimation error of $\theta_{i}$. In this case, the minimax criterion is modified to account for the relative importance of the parameters:
\begin{equation}
  \boldsymbol{\alpha}_{\text{opt}} = \underset{\boldsymbol{\alpha}}{\text{argmin}} \max \left[w_1\eta^{(N)}_{1}(\boldsymbol{\alpha}), w_{2} \eta^{(N)}_{2}(\boldsymbol{\alpha}), \cdots , w_{p} \eta^{(N)}_{p}(\boldsymbol{\alpha}) \right].  
\end{equation}

\section{Examples}\label{sec:applications}

In this section, we illustrate the proposed methodology by applying it to two-level quantum systems with ground state $\ket{0} = (0,1)^{T}$ and excited state $\ket{1} = (1,0)^{T}$. The corresponding Pauli matrices are defined as
\begin{equation}
\label{eq:Pauli}
 \sigma_{x}=\ket{0}\bra{1} + \ket{1}\bra{0}, \quad \sigma_{y}=i\ket{0}\bra{1}-i\ket{1}\bra{0}, \quad 
 \sigma_{z}=\ket{1}\bra{1}-\ket{0}\bra{0}.
\end{equation}
To avoid the prior distribution being overwhelmed by the measurement data, and to keep the analytical calculations computationally tractable, we restrict our attention to the estimation of two parameters with cost matrix $C=I_{2}$ and a small number of measurement outcomes. In the following examples, we adopt the standard quantum parameter estimation protocol, which consists of four stages: (i) state preparation, (ii) parameter encoding via a quantum operation, (iii) measurement, and (iv) estimation \cite{Pezze2018}.

\subsection{Estimation of phase rotations}
One of the most widely studied problems in quantum estimation is phase estimation. In this subsection, we demonstrate how the methodology presented in Sec.\ref{sec:nouveltechnique} can be applied to estimate phase rotations on the Bloch sphere. Recall that the quantum state of a two-level system can be represented geometrically by a three-dimensional Bloch vector $\mathbf{a}$, satisfying $\| \mathbf{a} \| \leq 1$. Quantum operations on the quantum state can then be visualized as changes in the length and orientation of the Bloch vector.
\begin{figure}[h]
    \centering
    \includegraphics[width=0.8\linewidth]{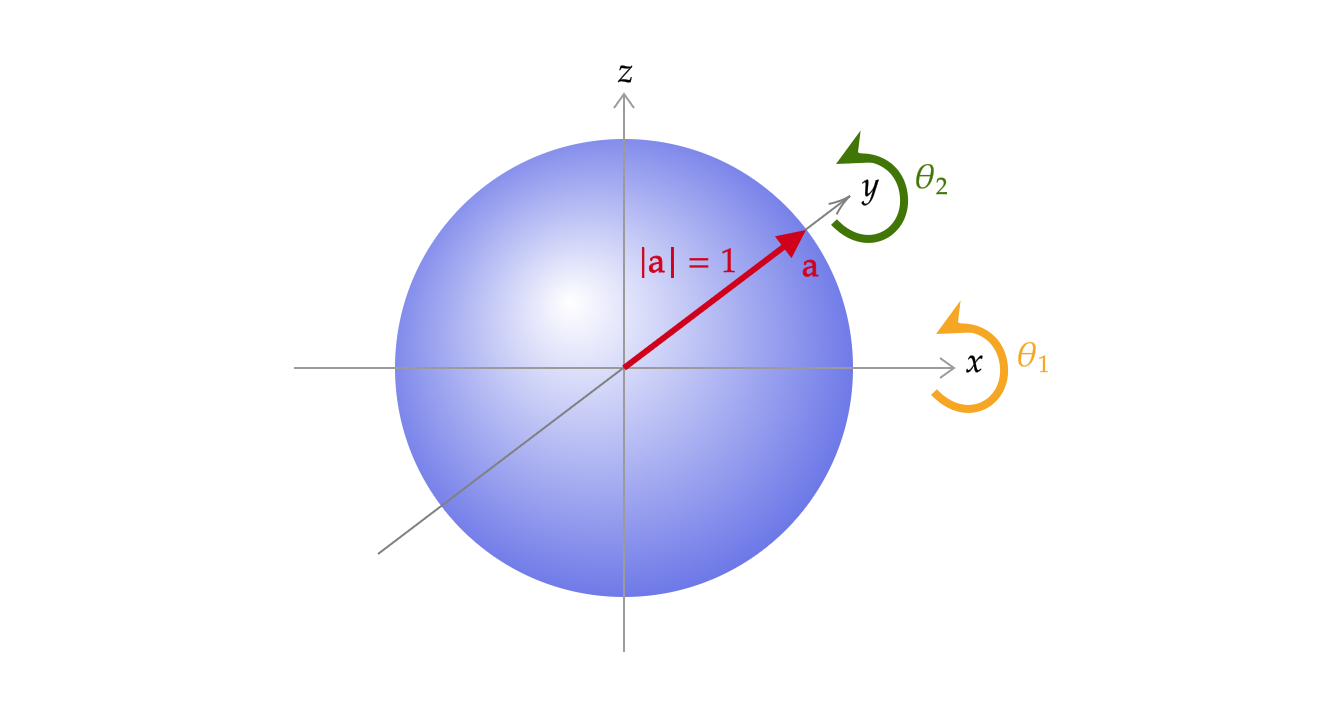}
    \caption{Schematic representation of the parameter-encoding process, or equivalently the realization of the quantum parametric model $\rho_{\boldsymbol{\theta}}$ with $\boldsymbol{\theta}=(\theta_1, \theta_2)^T$. The red arrow, representing the initial state, is first rotated about the $x$-axis by the unknown angle $\theta_{1}$ and subsequently about the $y$-axis by the unknown angle $\theta_{2}$.}
    \label{fig:example2}
\end{figure}
We first prepare the initial pure state with Bloch vector $\mathbf{a}=(0,1,0)^T$, i.e., 
\begin{equation}
    \rho = \frac{1}{2}\left(I_{2} + \sigma_{y} \right).
\end{equation}
This state is represented by a unit Bloch vector pointing along the positive $y$-axis, as shown in Fig. \ref{fig:example2}. Next, the system undergoes successive rotations generated by $\sigma_{x}$ and $\sigma_{y}$. These operations rotate the Bloch vector first about the
$x$-axis by the unknown angle $\theta_{1}$ and subsequently about the $y$-axis by the unknown angle $\theta_{2}$. As a result, the unknown parameters $\theta_{1}$ and $\theta_{2}$ are encoded in the density matrix according to
\begin{equation}
\label{eq:ex1parametric}
    \rho_{\boldsymbol{\theta}} = e^{-i \theta_{2} \sigma_{y}} e^{-i \theta_{1} \sigma_{x}} \rho e^{i \theta_{1} \sigma_{x}} e^{i \theta_{2} \sigma_{y}}. 
\end{equation}
The specification of the prior PDF $\pi(\boldsymbol{\theta})$ is an essential ingredient of Bayesian estimation, as it quantifies the information available about the unknown parameters before any measurements are performed. Since the choice of prior depends on the assumptions made about the estimation problem, several approaches based on Bayesian principles have been proposed in the literature \cite{Bernardo1994}. In the present work, we assume that, apart from the lower and upper bounds, no prior information about the unknown rotation angles is available before the measurement process. Accordingly, we assign equal probability to all admissible values. For illustrative purposes, both angles are assumed to be uniformly distributed over the interval
$[0, \frac{2\pi}{3}]$ with equal probability. The normalization condition 
\begin{equation}
    \int_{\Theta} \pi(\boldsymbol{\theta}) \, d \boldsymbol{\theta}= \int^{\frac{2\pi}{3}}_0 d\theta_{1} \int^{\frac{2\pi}{3}}_0 d\theta_{2} \, \pi(\theta_{1}, \theta_{2})=1
\end{equation}
then yields
\begin{equation}
\pi(\theta_{1}, \theta_{2}) = \frac{9}{4\pi^{2}},  
\end{equation}
for $\theta_{1}, \theta_{2} \in [0, \frac{2 \pi}{3}]$, and $\pi(\theta_{1}, \theta_{2})=0$ otherwise.

We now turn to the Personick approach. Using Eqs.\eqref{eq: Gammas} and \eqref{eq:ex1parametric}, we obtain the matrices
\begin{eqnarray}
 \Gamma_{0}&=&\int_\Theta \rho_{\boldsymbol{\theta}} \pi (\boldsymbol{\theta})\, d\boldsymbol{\theta}=\begin{pmatrix}
    0.463 & 0.064+0.103i  \\
    0.064-0.103i & 0.537
    \end{pmatrix},  \\
 \Gamma_{1,1}&=&\int_\Theta \theta_1 \rho_{\boldsymbol{\theta}} \pi (\boldsymbol{\theta})\, d\boldsymbol{\theta}=\begin{pmatrix}
    0.508 & 0.026+0.306i \\
    0.026-0.306i & 0.539
    \end{pmatrix},  \\
 \Gamma_{1,2}&=&\int_\Theta \theta_2 \rho_{\boldsymbol{\theta}} \pi (\boldsymbol{\theta})\, d\boldsymbol{\theta}=\begin{pmatrix}
    0.414 & 0.026+0.108i \\
    0.026-0.108i & 0.633
    \end{pmatrix}. 
\end{eqnarray}
It can be verified that $\Gamma_0$ is positive definite, as both of its eigenvalues are strictly positive. Thus, the Lyapunov equations in Eq.~\eqref{Solution1} yield the solutions:
\begin{eqnarray}
    \Lambda_{\text{opt},1} &=2 \int_{0}^{\infty} e^{-\Gamma_{0} y} \Gamma_{1,1} e^{-\Gamma_{0} y} dy= 
    \begin{pmatrix}
        1.017 & -0.072 + 0.411i \\
        -0.072 - 0.411i & 0.933
    \end{pmatrix} \nonumber  \\
    \Lambda_{\text{opt},2} &= 2 \int_{0}^{\infty} e^{-\Gamma_{0} y} \Gamma_{1,2} e^{-\Gamma_{0} y} dy=
    \begin{pmatrix}
        0.906 & -0.082 \\
        -0.082 & 1.189 
    \end{pmatrix} .
\end{eqnarray}
A direct calculation shows that
\begin{equation}
 [\Lambda_{\text{opt},1}, \Lambda_{\text{opt},2}] \neq 0   
\end{equation}
which implies that the two matrices do not share a common eigenbasis. Therefore, there does not exist a single PVM that is simultaneously optimal for the estimation of both parameters. This is precisely the situation in which the methodology developed in Sec.~\ref{sec:nouveltechnique} becomes relevant. Therefore, we construct the $\alpha$-dependent PVM by first forming the convex combination of the two Personick solutions according to Eq.~\eqref{eqn:convex},
\begin{equation}
\label{eq:Mexample1}
    M_{\alpha}= \alpha\Lambda_{\text{opt},1}+(1-\alpha)\Lambda_{\text{opt},2}.
\end{equation}
The normalized eigenvectors of $M_{\alpha}$ then define the corresponding two-element PVM, $\{P_\alpha, Q_\alpha\}$, satisfying
$P_\alpha+Q_\alpha=I_2$. We assign the measurement outcomes $x=1$ and $x=0$ to the projectors $P_\alpha$ and $Q_\alpha$. The corresponding conditional probabilities are given by
\begin{equation}
    p_{\alpha}(x=1 \vert \boldsymbol{\theta}) = \Tr[\rho_{\boldsymbol{\theta}}P_{\alpha}] \quad \text{and} \quad
    p_{\alpha}(x=0 \vert \boldsymbol{\theta}) = \Tr[\rho_{\boldsymbol{\theta}}Q_{\alpha}]=1-p_{\alpha}(x=1 \vert \boldsymbol{\theta}).
\end{equation}

\begin{figure}[t!]
  \centering
  \begin{subfigure}[b]{0.49\textwidth}
    \centering
    \includegraphics[width=\textwidth]{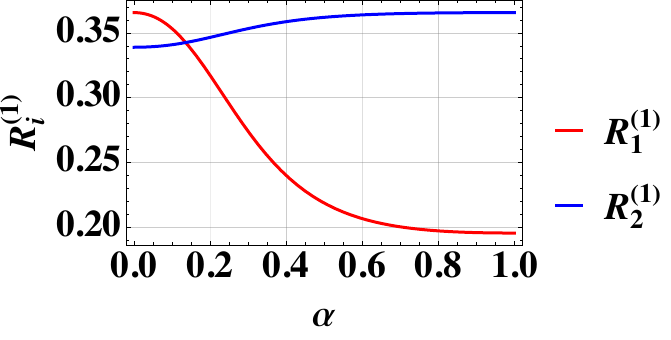}
    \caption{BMSE of $\theta_{1}$ and $\theta_{2}$}
    \label{fig:3a}
  \end{subfigure}
  \hfill
  \begin{subfigure}{0.49\textwidth}
    \centering
    \includegraphics[width=\textwidth]{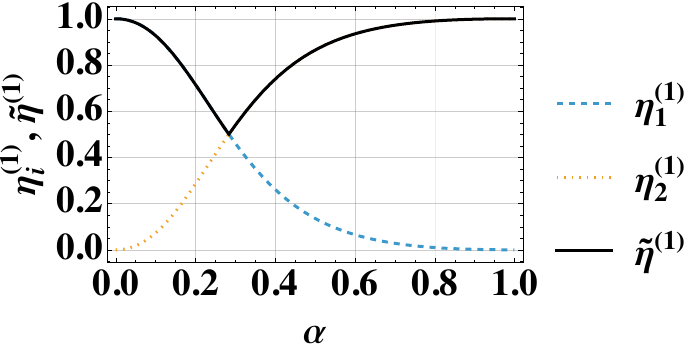}
    \caption{Rescaled BMSE of $\theta_{1}$ and $\theta_{2}$}
    \label{fig:3b}
  \end{subfigure}
  \caption{Bayesian mean-square errors (BMSEs) of the rotation angles $\theta_{1}$ and $\theta_{2}$ in the phase-rotation estimation problem. (a) BMSEs as functions of the variational parameter $\alpha$. (b) Corresponding min-max normalized BMSEs together with $\tilde{\eta}^{(1)}=\max\left(\eta^{(1)}_{1},\eta^{(1)}_{2}\right)$, whose minimum determines $\alpha_{ \text{opt}}$.}
  \label{fig:3}
\end{figure}

The next stage is the measurement phase, in which an $N$-point dataset
$\mathbf{x}=\{x_1, x_2, \cdots, x_N\}$, with $x_i\in\{0,1\}$, is collected. From these measurement outcomes, the likelihood function in Eq.~\eqref{eq: likelihoodf} is constructed, the MMSE estimators in Eq.~\eqref{eq:MMSEcl} are computed using the posterior PDF in Eq.~\eqref{eq:posteriorcl}, and finally the corresponding BMSEs of $\theta_{1}$ and $\theta_{2}$ are evaluated according to Eq.~\eqref{eqn:TotalBMSE}. For simplicity, while still illustrating the proposed methodology, we consider the case of a single measurement outcome, i.e., $N=1$. The resulting BMSEs of $\theta_{1}$ and $\theta_{2}$ are shown in Fig. \ref{fig:3}. Figure~\ref{fig:3a} shows that the BMSE of $\theta_{1}$ is smallest at $\alpha=1$, whereas that of $\theta_{2}$ is smallest at $\alpha=0$, as expected from the convex combination in Eq. \eqref{eq:Mexample1}. It is also evident that the BMSEs cannot be compared directly, since the variation of $\mathcal{R}^{(1)}_{1}$ over the interval $\alpha \in [0,1]$ is significantly larger than that of $\mathcal{R}^{(1)}_{2}$. To place the two quantities on an equal footing, we therefore rescale the BMSEs using the min-max normalization introduced in Eq. \eqref{eq:normalization}. The normalized BMSEs are shown as dotted curves in Fig.~\ref{fig:3b}. The optimized value of $\alpha$ is determined by minimizing
\begin{equation}
\tilde{\eta}^{(1)}(\alpha)=\max\left(\eta^{(1)}_{1}(\alpha),\eta^{(1)}_{2}(\alpha)\right),    
\end{equation}
which, in this example, occurs at the intersection of the two normalized BMSE curves, i.e.,
\begin{equation}
  \eta^{(1)}_{1}(\alpha_{ \text{opt}})=\eta^{(1)}_{2}(\alpha_{ \text{opt}}).  
\end{equation}
This yields $\alpha_{\text{opt}} \approx 0.284$, corresponding to the projective measurements
\begin{equation}
    P_{\alpha_{\text{opt}}} =
    \begin{pmatrix}
        0.768 & 0.237-0.349i \\
        0.237+0.349i & 0.232
    \end{pmatrix} \quad \text{and} \quad Q_{\alpha_{\text{opt}}}=I_2-P_{\alpha_{\text{opt}}}.
\end{equation}

\begin{figure}[t!]
    \centering
    \includegraphics[width=0.6\linewidth]{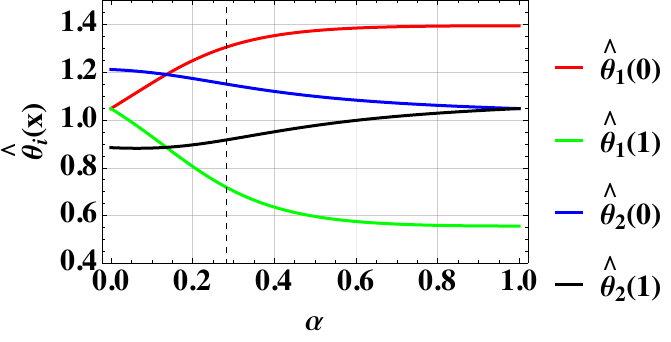}
    \caption{MMSE estimators in the phase-rotation estimation problem as functions of the variational parameter $\alpha$ for the different measurement outcomes obtained from a single measurement. The dashed line indicates the optimized value $\alpha_{\text{opt}}$ selected by the proposed approach.}
    \label{fig:estimators_phase}
\end{figure}

The MMSE estimates of $\theta_{1}$ and  $\theta_{2}$ for the different measurement outcomes $x \in \{0,1\}$ as functions of $\alpha$ are displayed in Fig.~\ref{fig:estimators_phase}. It can be observed that when $\alpha$ is chosen such that the measurement is optimized for the estimation of one rotation angle, the MMSE estimate of the other one becomes independent of the measurement outcome. Specifically, when $\alpha=0$, we have $\hat{\theta}_1(0)=\hat{\theta}_1(1)$, whereas for
$\alpha=1$, $\hat{\theta}_2(0)=\hat{\theta}_2(1)$. In these cases, the estimate depends only on the prior PDF $\pi(\theta_1, \theta_2)$. Thus, the MMSE estimate is equal to the mean of the uniform prior over the interval $[0, \frac{2\pi}{3}]$,
namely $\pi/3 \approx 1.047$. 

The BMSEs achieved by the proposed strategy  are $\mathcal{R}^{(1)}_{1}(\alpha_{\text{opt}})=0.280$ and $\mathcal{R}^{(1)}_{2}(\alpha_{\text{opt}})=0.352$. For comparison, when $\alpha=0$, the corresponding BMSEs are $\mathcal{R}^{(1)}_{1}(0)=0.366$ and $\mathcal{R}^{(1)}_{2}(0)=0.339$. In contrast, for $\alpha=1$, they are $\mathcal{R}^{(1)}_{1}(1)=0.195$ and $\mathcal{R}^{(1)}_{2}(1)=0.366$. These results illustrate precisely how the proposed method operates. A single quantum measurement scheme is employed to estimate both unknown rotation angles from the same dataset. As expected, the resulting BMSEs represent a balanced compromise, lying between the extreme BMSE values attained at $\alpha=0$ and $\alpha=1$.

\subsection{Estimation of parameters in convex combinations of unitary operations}

Our next example considers a two-level system prepared in the pure state with Bloch vector $\mathbf{a}=(1,0,0)^T$, i.e., 
\begin{equation}
    \rho = \frac{1}{2} \left(I_{2} + \sigma_{x} \right) .
\end{equation}
The parameter-encoding stage is inspired by the random unitary transformation introduced in \cite{Novotny2009}, which has applications in quantum networks. We fix three unitary matrices, namely the Pauli matrix $\sigma_{x}$ and $\sigma_{z}$, together with the Hadamard gate $H=\frac{1}{\sqrt{2}}(\sigma_{x}+\sigma_{z})$, and use the mixing probabilities $\theta_{1}$ and $\theta_{2}$ as the unknown parameters to be estimated. The quantum parametric model is given by the convex combination of unitary transformations
\begin{eqnarray}
\label{eq:ex2parametric}
    \rho_{\boldsymbol{\theta}} &= \theta_{1} \sigma_{x} \rho \sigma_{x} + \theta_{2} H \rho H + (1-\theta_{1}-\theta_{2}) \sigma_{z} \rho \sigma_{z},
\end{eqnarray} 
where the admissible parameter region $\Theta$ is
\begin{equation}
    0 \leqslant \theta_1, \quad 0 \leqslant \theta_2, \quad \theta_1+\theta_2 \leqslant 1. 
\end{equation}
We assume that no prior information about $\theta_{1}$ and $\theta_{2}$ is available and therefore assign a uniform prior over the admissible parameter region. The normalization condition
\begin{equation}
    \int_{\Theta} \pi(\boldsymbol{\theta}) \, d \boldsymbol{\theta}= \int^1_0 d\theta_{1} \int^{1-\theta_1}_0 d\theta_{2} \, \pi(\theta_{1}, \theta_{2})=1
\end{equation}
yields
\begin{equation}
\pi(\theta_{1}, \theta_{2}) = 2  
\end{equation}
for $(\theta_1,\theta_2)\in \Theta$, and $\pi(\theta_1,\theta_2)=0$ otherwise. 

Following the same procedure as in the previous example, we apply the Personick formalism to the present quantum parametric model. Evaluating the matrices defined in Eqs.~\eqref{eq: Gammas} and \eqref{eq:ex2parametric} gives
\begin{eqnarray}
 \Gamma_{0}&=&\int_\Theta \rho_{\boldsymbol{\theta}} \pi (\boldsymbol{\theta})\, d\boldsymbol{\theta}=\begin{pmatrix}
        \frac{2}{3} & 0 \\
        0 & \frac{1}{3}
    \end{pmatrix},  \\
 \Gamma_{1,1}&=&\int_\Theta \theta_1 \rho_{\boldsymbol{\theta}} \pi (\boldsymbol{\theta})\, d\boldsymbol{\theta}=\begin{pmatrix}
        \frac{5}{24} & \frac{1}{24} \\
        \frac{1}{24} & \frac{1}{8}
    \end{pmatrix},  \\
 \Gamma_{1,2}&=&\int_\Theta \theta_2 \rho_{\boldsymbol{\theta}} \pi (\boldsymbol{\theta})\, d\boldsymbol{\theta}=\begin{pmatrix}
        \frac{1}{4} & 0 \\
        0 & \frac{1}{12}
    \end{pmatrix}. 
\end{eqnarray}
Since $\Gamma_0$ is positive definite, the Lyapunov equations in Eq.~\eqref{Solution1} admit a unique solution, yielding
\begin{eqnarray}
    \Lambda_{\text{opt},1} &=2 \int_{0}^{\infty} e^{-\Gamma_{0} y} \Gamma_{1,1} e^{-\Gamma_{0} y} dy= 
    \begin{pmatrix}
        \frac{5}{16} & \frac{1}{12} \\
        \frac{1}{12} & \frac{3}{8}
    \end{pmatrix} \nonumber  \\
    \Lambda_{\text{opt},2} &= 2 \int_{0}^{\infty} e^{-\Gamma_{0} y} \Gamma_{1,2} e^{-\Gamma_{0} y} dy=
    \begin{pmatrix}
        \frac{3}{8} & 0 \\
        0 & \frac{1}{4}
    \end{pmatrix} .
\end{eqnarray}
As in the previous example, the Personick solutions do not commute, i.e.,
\begin{equation}
 [\Lambda_{\text{opt},1}, \Lambda_{\text{opt},2}] \neq 0,   
\end{equation}
and therefore no common eigenbasis exists. Thus, no single projective measurement is simultaneously optimal for estimating both parameters. Following the procedure described in Sec.~\ref{sec:nouveltechnique}, we construct the matrix $M_{\alpha}$ according to Eq.~\eqref{eqn:convex}. Its spectral decomposition yields the projectors 
\begin{equation}
    P_{\alpha} = 
    \begin{pmatrix}
        \frac{9\alpha-6+\sqrt{36+\alpha(145\alpha-108)}}{2\sqrt{36+\alpha(145\alpha-108)}} & -\frac{4\alpha}{\sqrt{36+\alpha(145\alpha-108)}} \\
        -\frac{4\alpha}{\sqrt{36+\alpha(145\alpha-108)}} & \frac{64\alpha^{2}}{64\alpha^{2}+\left(9\alpha-6+\sqrt{36+\alpha(145\alpha-108)}\right)^{2}}
    \end{pmatrix} \quad \text{and} \quad Q_{\alpha}=I_2-P_{\alpha}.
\end{equation}
As before, we assign the measurement outcomes $x=1$ and $x=0$ to the projectors $P_\alpha$ and $Q_\alpha$, respectively. The corresponding conditional probabilities are
\begin{equation}
    p_{\alpha}(x=1 \vert \boldsymbol{\theta}) = \Tr[\rho_{\boldsymbol{\theta}}P_{\alpha}] \quad \text{and} \quad
    p_{\alpha}(x=0 \vert \boldsymbol{\theta}) = \Tr[\rho_{\boldsymbol{\theta}}Q_{\alpha}]=1-p_{\alpha}(x=1 \vert \boldsymbol{\theta}).
\end{equation}
\begin{figure}[t!]
  \centering
  \begin{subfigure}[b]{0.49\textwidth}
    \centering
    \includegraphics[width=\textwidth]{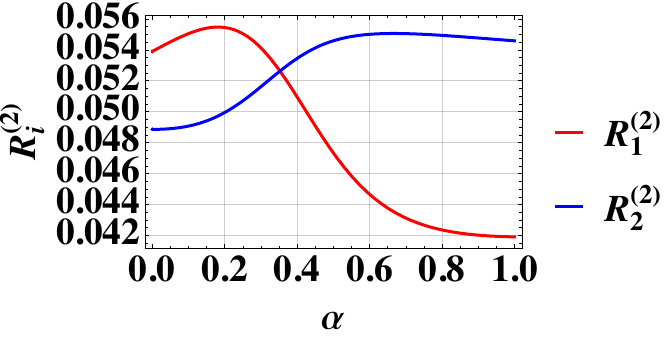}
    \caption{BMSE of $\theta_{1}$ and $\theta_{2}$}
    \label{fig:1a}
  \end{subfigure}
  \hfill
  \begin{subfigure}[b]{0.49\textwidth}
    \centering
    \includegraphics[width=\textwidth]{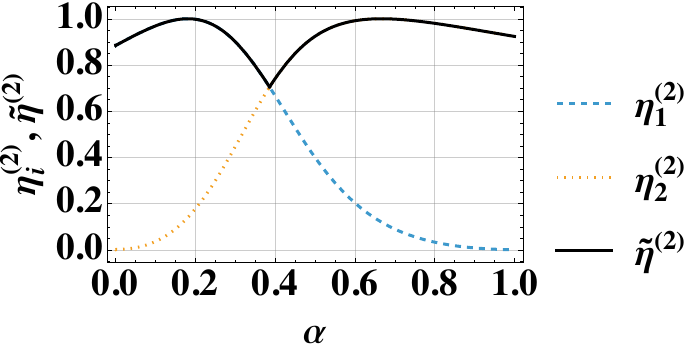}
    \caption{Rescaled BMSE of $\theta_{1}$ and $\theta_{2}$}
        \label{fig:1b}
  \end{subfigure}
  \caption{Bayesian mean-square errors (BMSEs) of mixing probabilities $\theta_{1}$ and $\theta_{2}$ in the convex combination of the unitary transformation. (a) BMSEs as functions of the variational parameter $\alpha$. (b) Corresponding min-max normalized BMSEs together with $\tilde{\eta}^{(2)}=\max\left(\eta^{(2)}_{1},\eta^{(2)}_{2}\right)$, whose minimum determines $\alpha_{ \text{opt}}$.}
  \label{fig:1}
\end{figure}

The measurement stage and the subsequent classical post-processing are carried out as described in the previous example. For simplicity, we consider the case of two measurement outcomes, i.e., $N=2$, from which the MMSE estimators and the corresponding BMSEs of $\theta_1$ and $\theta_2$ are computed. The results are shown in Fig.~\ref{fig:1}. Figure~\ref{fig:1a} displays the BMSEs of the two parameters as functions of $\alpha$. As can be observed, the BMSE of $\theta_{1}$ first increases to its local maximum before decreasing to the global minimum at $\alpha=1$, while that of $\theta_{2}$ increases from its global minimum to a local maximum and decreases slightly as $\alpha$ varies from $0$ to $1$. Another noteworthy observation is that the range of $\mathcal{R}_{1}^{(2)}$, which spans approximately from $0.042$ to $0.055$, is this time only slightly larger than that of $\mathcal{R}_{2}^{(2)}$, which ranges from $0.049$ to $0.055$. The dotted curves in Fig.~\ref{fig:1b} show the corresponding min-max normalized BMSEs. As in the previous example, the optimized value $\alpha_{\text{opt}}$ is determined by the minimum of
\begin{equation}
    \tilde{\eta}^{(2)}(\alpha) = \max\left(\eta_{1}^{(2)}(\alpha), \eta_{2}^{(2)}(\alpha)\right),
\end{equation}
which, by construction, coincides with the intersection of the two normalized BMSE curves. In this example, we obtain $\alpha_{\text{opt}}=0.386$. The resulting projective measurements are given by
\begin{equation}
    P_{\alpha_{\text{opt}}} =
    \begin{pmatrix}
        0.184 & -0.387 \\
        -0.387 & 0.816
    \end{pmatrix} \quad \text{and} \quad Q_{\alpha_{\text{opt}}}=I_2-P_{\alpha_{\text{opt}}}.
\end{equation}

\begin{figure}[t!]
    \centering
    \includegraphics[width=0.7\linewidth]{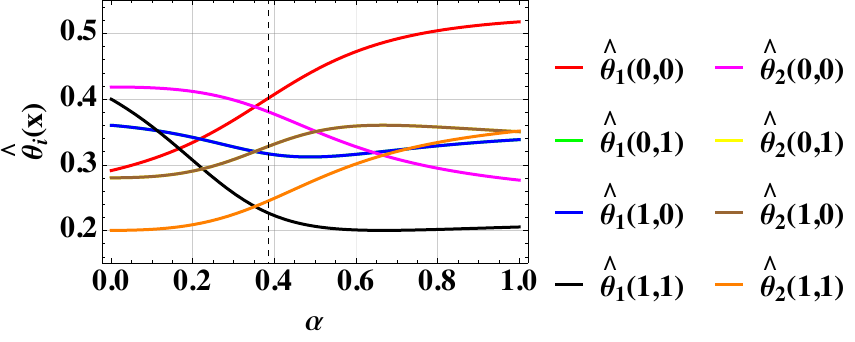}
    \caption{MMSE estimators in the estimation of the mixing probabilities in random unitary transformations as functions of the variational parameter $\alpha$ for each possible measurement outcome obtained from two measurements. The dashed line indicates the optimized value $\alpha_{\text{opt}}$ selected by the proposed methodology. The measurement outcomes $\mathbf{x}=\{0,1\}$ and $\mathbf{x}=\{1,0\}$ lead to identical estimates, namely $\hat{\theta_1}(0,1)=\hat{\theta_1}(1,0)$ and $\hat{\theta_2}(0,1)=\hat{\theta_2}(1,0)$, resulting in only six distinct curves (rather than eight) in the plot.}
    \label{fig:estimators_RUT}
\end{figure}

Figure~\ref{fig:estimators_RUT} shows the MMSE estimators of $\theta_{1}$ and $\theta_{2}$ for the different measurement outcomes $\{0,0\}$, $\{0,1\}$, $\{1,0\}$, $\{1,1\}$ as a function of the variational parameter $\alpha$. One interesting observation is that the measurement outcomes $\mathbf{x}=\{0,1\}$ and $\mathbf{x}=\{1,0\}$ lead to identical (degenerate) estimates for both $\theta_{1}$ and $\theta_{2}$ for all values of $\alpha \in [0,1]$. We can therefore infer that the order of the measurement outcomes plays no role in the estimation process when two measurements are performed. This is consistent with the general principle that the inference is determined usually by the observed measurement statistics, rather than by the sequence in which the outcomes are recorded. The optimized value, $\alpha=\alpha_{\text{opt}}$, is indicated by the vertical dashed line.
At this point, the corresponding BMSEs are $\mathcal{R}_{1}^{(2)} (\alpha_{\text{opt}}) = 0.051$ and $\mathcal{R}_{2}^{(2)}(\alpha_{\text{opt}}) = 0.053$ respectively. In comparison, the BMSEs in the case $\alpha=0$ are $\mathcal{R}_{1}^{(2)}(0)=0.054$ and $\mathcal{R}_{2}^{(2)}(0)=0.049$ respectively, whereas their corresponding values for $\alpha=1$ are $\mathcal{R}_{1}^{(2)}(1)=0.042$ and $\mathcal{R}_{2}^{(2)}(1)=0.055$, respectively. As expected, the proposed strategy yields a compromise between these two extreme cases, providing a single projective measurement with balanced estimation performance for both parameters.

\section{Concluding remarks}
\label{sec:conluding}

In this work, we have developed a general framework for Bayesian multiparameter estimation based on the approach introduced by Personick \cite{Personick1970, Personick1971}. Already in his 1970 technical report \cite{Personick1970}, Personick recognized that the optimal observables associated with different parameters generally do not commute and proposed a Lagrange multiplier formulation for the multiparameter problem, although no general solution was obtained. Our objective was therefore to construct a single projection-valued measure (PVM) from which all unknown parameters can be inferred. This is particularly important in the Bayesian framework because, once a fixed PVM has been specified, the corresponding likelihood function is determined, and the minimum mean-square error (MMSE) estimator associated with each parameter minimizes its corresponding Bayesian mean-square error (BMSE) \cite{Kay2009}. To address this challenge, we construct a convex combination of the solutions to the Lyapunov equations arising in Personick's formalism. This construction introduces a set of variational parameters satisfying the constraints of the convex combination. For a problem involving $p$ unknown parameters, the number of variational parameters is $p-1$. The spectral decomposition of the operator resulting from the convex combination defines a parameterized PVM. Consequently, both the BMSEs and the corresponding MMSE estimators become functions of the variational parameters. Since the BMSEs may differ substantially in magnitude, for example, when simultaneously estimating the dimensionless temperature of a quantum system \cite{Brunelli2011} or a high frequency \cite{Bernad2019}, and thus their direct comparison is not meaningful. To overcome this difficulty, we introduce a min–max normalization procedure that places the BMSEs on a common scale and enables a meaningful minimax optimization. The variational parameters are then determined by minimizing the maximum of the normalized BMSEs, thereby yielding a variationally optimized PVM that overcomes the fundamental obstacle posed by the noncommutativity of the optimal observables. Although global optimality is not guaranteed, the proposed variational optimization defines a well-posed strategy for constructing optimized PVMs.

To illustrate the proposed method, we considered two representative qubit estimation problems: the estimation of phase rotations and the estimation of parameters defining convex combinations of unitary operations. These proof-of-principle examples were analyzed for two unknown parameters, uniform prior probability density functions, and small data sets consisting of one and two measurement outcomes. In both cases, the proposed optimization procedure successfully identified a single PVM from which the MMSE estimators of all unknown parameters were obtained. The approach is nevertheless accompanied by computational challenges. The eigendecomposition of the operator obtained from the convex combination of the Lyapunov solutions often yields highly involved analytical expressions for the corresponding eigenprojectors, making their symbolic manipulation and the subsequent evaluation of the BMSEs increasingly demanding. These calculations remain tractable for the examples considered here, but the complexity of the symbolic expressions grows rapidly with the size of the data set. Numerical evaluation remains feasible, although the intermediate arrays required for the evaluation of the BMSEs, and hence the number of computational steps, increase rapidly with the size of the data set.

In conclusion, the framework developed here extends Personick's Bayesian estimation theory to the general multiparameter setting by providing a constructive procedure for designing projective measurements. Moreover, the proposed variational optimization and min–max normalization techniques may prove useful in broader quantum estimation and optimization problems involving multiple competing performance criteria.

\section*{Acknowledgement}

J Z B was supported by the Hungarian National Research, Development and Innovation Office within the Quantum Information National Laboratory of Hungary grants no. 2022-2.1.1-NL-2022-00004 and 134437.

\appendix
\section{Derivation of the Lyapunov equations for the general N-copy case}
\label{sec:AppendixA}
In the main text, we consider only the single-copy setting. In this appendix, we provide a detailed derivation of the Lyapunov equation presented in Sec.~\ref{sec:preliminaries} and generalize it to the N-copy setting introduced in Eq.~\eqref{eq:Ntensor}. We use the notation introduced in Sec.~\ref{sec:preliminaries}. As in the main text, the $N$ measurement outcomes are collected in the vector $\mathbf{x}\in \{x_1,x_2,\dots,x_N\}$. We employ the quadratic cost function introduced in the main text, $\mathcal{C}\big[\hat{\boldsymbol{\theta}}
(\mathbf{x}), \boldsymbol{\theta}\big] =\left(\hat{\boldsymbol{\theta}}(\mathbf{x})-\boldsymbol{\theta} \right)^{T} C \left( \hat{\boldsymbol{\theta}}(\mathbf{x})-\boldsymbol{\theta} \right)$, where $C$, $\boldsymbol{\theta}$ and $\hat{\boldsymbol{\theta}}(\mathbf{x})$ denote the cost matrix, the parameters of interest and their estimators, respectively. The derivation starts from the BMSE, given by
\begin{eqnarray}
    &\mathcal{R}^{(N)} = \sum_{\mathbf{x}\in \mathcal{X}^{N}} \int_{\Theta}   \left(\hat{\boldsymbol{\theta}}(\mathbf{x})-\boldsymbol{\theta} \right)^{T} C \left( \hat{\boldsymbol{\theta}}(\mathbf{x})-\boldsymbol{\theta} \right) \Tr\left[\rho_{\boldsymbol{\theta}}^{\otimes N} \Pi_{\mathbf{x}}^{(N)}\right] \pi(\boldsymbol{\theta}) \, d\boldsymbol{\theta} \\
    &= \sum_{\mathbf{x}\in \mathcal{X}^{N}} \int_{\Theta} \left( \hat{\boldsymbol{\theta}}^{T}(\mathbf{x})C\hat{\boldsymbol{\theta}}(\mathbf{x})-2 \boldsymbol{\theta}^{T}  C\hat{\boldsymbol{\theta}}(\mathbf{x}) +\boldsymbol{\theta}^{T}C\boldsymbol{\theta} \right)  \Tr\left[\rho_{\boldsymbol{\theta}}^{\otimes N} \Pi_{\mathbf{x}}^{(N)}\right] \pi(\boldsymbol{\theta}) \, d\boldsymbol{\theta}, \nonumber
\end{eqnarray}
where $\pi(\boldsymbol{\theta})$ is the prior PDF. The notation $\mathcal{X}^{N}$ denotes the set of all possible measurement outcomes obtained from $N$ measurements. Note that  $\boldsymbol{\theta}$, $\hat{\boldsymbol{\theta}}(\mathbf{x})$, and $C$ belong to the parameter space, whereas the positive semidefinite operator $\Pi_{\mathbf{x}}^{(N)}$ and the density operator $\rho_{\boldsymbol{\theta}}^{\otimes N}$ act on the Hilbert space $\mathcal{H}^{\otimes N}$. Since operators acting on different spaces commute, we can rewrite the above expression as
\begin{eqnarray}
    &\mathcal{R}^{(N)} = \int_{\Theta} d\boldsymbol{\theta} \Tr \left[\rho_{\boldsymbol{\theta}}^{\otimes N} \pi(\boldsymbol{\theta}) \,\sum_{\mathbf{x}\in \mathcal{X}^{N}} \hat{\boldsymbol{\theta}}^{T}(\mathbf{x})C\hat{\boldsymbol{\theta}}(\mathbf{x}) \Pi_{\mathbf{x}}^{(N)}\right] \\
    & - 2 \int_{\Theta} d\boldsymbol{\theta}\sum_{\mathbf{x} \in \mathcal{X}^{N}} \Tr\left[\boldsymbol{\theta}^{T}  C\hat{\boldsymbol{\theta}}(\mathbf{x})\rho_{\boldsymbol{\theta}}^{\otimes N}\pi(\boldsymbol{\theta}) \Pi_{\mathbf{x}}^{(N)} \right] \nonumber \\
    &+ \int_{\Theta} d\boldsymbol{\theta} \,\boldsymbol{\theta}^{T} C \boldsymbol{\theta} \pi(\boldsymbol{\theta}) \Tr \left[\rho_{\boldsymbol{\theta}}^{\otimes N}  \sum_{\mathbf{x}\in \mathcal{X}^{N}} \Pi_{\boldsymbol{x}}^{(N)} \right]. \nonumber
\end{eqnarray}
In analogy with the single-copy case, we define the quantities
\begin{eqnarray}
\boldsymbol{\Lambda}_{1}^{(N)} &= \sum_{\mathbf{x}\in \mathcal{X}^{N}} \hat{\boldsymbol{\theta}}(\boldsymbol{x}) \Pi_{\boldsymbol{x}}^{(N)} , \qquad \Lambda_{2}^{(N)}=\sum_{\mathbf{x}\in \mathcal{X}^{N}} \hat{\boldsymbol{\theta}}^{T}(\boldsymbol{x}) C \hat{\boldsymbol{\theta}} \Pi_{\boldsymbol{x}}^{(N)}, \\
\Gamma_{0}^{(N)} &= \int_{\Theta} d\boldsymbol{\theta} \rho_{\boldsymbol{\theta}}^{\otimes N} \pi(\boldsymbol{\theta}), \quad 
\boldsymbol{\Gamma}_{1}^{(N)} =\int_{\Theta} d\boldsymbol{\theta} \,\boldsymbol{\theta} \,\rho_{\boldsymbol{\theta}}^{\otimes N} \pi(\boldsymbol{\theta}), \nonumber \\
\Delta_{C}^{2} \boldsymbol{\theta} &= \int_{\Theta} d\boldsymbol{\theta} \,\boldsymbol{\theta}^{T} C \boldsymbol{\theta} \,\pi(\boldsymbol{\theta}). \nonumber
\end{eqnarray}
Using the identities $\Tr\left[\rho_{\boldsymbol{\theta}}^{\otimes N}\right]=1$ and $\sum_{\mathbf{x}\in \mathcal{X}^{N}} \Pi_{\mathbf{x}}^{(N)} = I^{(N)}$, where $I^{(N)}$ denotes the identity operator on the Hilbert space $\mathcal{H}^{\otimes N}$, the BMSE can be written as
\begin{equation}
    \mathcal{R}^{(N)} = \Tr \left[ \Gamma_{0}^{(N)} \Lambda_{2}^{(N)}\right] - 2 \Tr\left[\left(\boldsymbol{\Gamma}_{1}^{(N)}\right)^T C \boldsymbol{\Lambda}_{1}^{(N)}\right] + \Delta_{C}^{2} \boldsymbol{\theta}
\end{equation}

Since $ \Pi_{\mathbf{x}}^{(N)} \geq 0$ and the quadratic form $\left(\hat{\boldsymbol{\theta}}(\boldsymbol{x})-\boldsymbol{\Lambda}_{1}^{(N)}\right)^{T} C \left(\hat{\boldsymbol{\theta}}(\boldsymbol{x})-\boldsymbol{\Lambda}_{1}^{(N)}\right) \geq 0$, their product is positive semidefinite. Thus,
\begin{equation}
 \sum_{\mathbf{x}\in \mathcal{X}^{N}} \Pi_{\mathbf{x}}^{(N)}    \left(\hat{\boldsymbol{\theta}}(\boldsymbol{x})-\boldsymbol{\Lambda}_{1}^{(N)}\right)^{T} C \left(\hat{\boldsymbol{\theta}}(\boldsymbol{x})-\boldsymbol{\Lambda}_{1}^{(N)}\right) \geq 0.
\end{equation}
Expanding the quadratic form and using the relation $\sum_{\mathbf{x}\in \mathcal{X}^{N}} \Pi_{\mathbf{x}}^{(N)} = I^{(N)}$, the cross terms combine according to the definition of $\boldsymbol{\Lambda}_{1}^{(N)}$, while the quadratic term yields $\Lambda_{2}^{(N)}$. Consequently,
$\Lambda_{2}^{(N)} \geq \left(\boldsymbol{\Lambda}_{1}^{(N)}\right)^T C \boldsymbol{\Lambda}_{1}^{(N)}$. The BMSE is therefore bounded below by
\begin{equation}
\label{eq:appendixn}
    \mathcal{R}^{(N)} \geq B(\boldsymbol{\Lambda}_{1}^{(N)}):= \Delta_{C}^{2} \boldsymbol{\theta} - 2 \Tr \left[ \left(\boldsymbol{\Gamma}_{1}^{(N)}\right)^T C \boldsymbol{\Lambda}_{1}^{(N)} \right] + \Tr\left[\Gamma_{0}^{(N)}\left(\boldsymbol{\Lambda}_{1}^{(N)}\right)^T C \boldsymbol{\Lambda}_{1}^{(N)}\right]
\end{equation}
Since $\Gamma_{0}^{(N)}$ is a scalar with respect to the parameter space $\Theta$ and $\left(\boldsymbol{\Lambda}_{1}^{(N)}\right)^T C \boldsymbol{\Lambda}_{1}^{(N)}$ is a scalar quadratic form, we may use the cyclic property of the parameter-space trace $\tr[\cdot]$ to write
\begin{equation}
    B(\boldsymbol{\Lambda}_{1}^{(N)})=\Delta_{C}^{2} \boldsymbol{\theta} + \tr\left[ C \left\{ \Tr\left[\Gamma_{0}^{(N)}\boldsymbol{\Lambda}_{1}^{(N)} \left(\boldsymbol{\Lambda}_{1}^{(N)}\right)^T\right] - 2 \Tr \left[ \boldsymbol{\Lambda}_{1}^{(N)}\left(\boldsymbol{\Gamma}_{1}^{(N)}\right)^T \right] \right\} \right].
\end{equation}
We now return to Eq.~\eqref{eq:appendixn} and determine $\boldsymbol{\Lambda}^{(N)}_{\text{opt}}$, which minimizes 
$B(\boldsymbol{\Lambda}_{1}^{(N)})$, following the variational approach introduced by Personick \cite{Personick1971}. Let $\epsilon \in \mathbb{R}$ and $\boldsymbol{H}$ be an arbitrary $p$-dimensional vector whose entries are self-adjoint operators acting on the Hilbert space $\mathcal{H}^{\otimes N}$. Then
\begin{equation}
    B(\boldsymbol{\Lambda}^{(N)}_{\text{opt}}) \leq B(\boldsymbol{\Lambda}^{(N)}_{\text{opt}}+\epsilon \boldsymbol{H})
\end{equation}
since $\boldsymbol{\Lambda}^{(N)}_{\text{opt}}+\epsilon \boldsymbol{H}$ is also a vector of self-adjoint operators. Expanding the right-hand side of the inequality and differentiating with respect to $\epsilon$ at $\epsilon=0$ yields the first-order optimality condition
\begin{equation}
    \Tr\left[ \boldsymbol{H}^T C \left(\boldsymbol{\Lambda}_{\text{opt}}^{(N)} \Gamma_{0}^{(N)} + \Gamma_{0}^{(N)}\boldsymbol{\Lambda}_{ \text{opt}}^{(N)} - 2\boldsymbol{\Gamma}_{1}^{(N)}\right)\right]=0,
\end{equation}
which must hold for every admissible $\boldsymbol{H}$. Since $C$ is positive definite, it is invertible. Hence, the vector $\boldsymbol{H}^T C$ is arbitrary whenever $\boldsymbol{H}^T$ is arbitrary. Therefore, the above identity holds for every admissible operator-valued vector $\boldsymbol{H}^T C$, which is possible only if
\begin{equation}
    \boldsymbol{\Lambda}_{\text{opt}}^{(N)} \Gamma_{0}^{(N)} + \Gamma_{0}^{(N)}\boldsymbol{\Lambda}_{ \text{opt}}^{(N)} = 2\boldsymbol{\Gamma}_{1}^{(N)}.
\end{equation}
If $\Gamma_{0}^{(N)}>0$, the Lyapunov equations admit a unique self-adjoint solution, given by \cite{Personick1971} 
\begin{equation}\label{Solution}
    \boldsymbol{\Lambda}_{\text{opt}}^{(N)} = 2 \int_{0}^{\infty} e^{-\Gamma_{0}^{(N)} x} \boldsymbol{\Gamma}_{1}^{(N)} e^{-\Gamma_{0}^{(N)} x} dx.
\end{equation}

Our first observation is that the form of the equations is identical in the single-copy and the $N$-copy settings. Moreover, the operators $\Pi_{\mathbf{x}}^{(N)}$ entering the construction of $\boldsymbol{\Lambda}_{1}^{(N)}$ are arbitrary positive semidefinite operators acting on the Hilbert space $\mathcal{H}^{\otimes N}$. Therefore, it is not immediately clear whether the solution $\boldsymbol{\Lambda}_{\text{opt}}^{(N)}$, and hence the projectors appearing in its spectral decomposition, possesses any particular structure. In view of the tensor-product structure of the state $\rho_{\boldsymbol{\theta}}^{\otimes N}$, one may ask whether these projectors are of the form $P_1\otimes P_2\otimes \dots \otimes P_N$, where each $P_i$ is a projector acting on the corresponding single-copy Hilbert space $\mathcal{H}$. To address this question, we consider the estimation of a single parameter.

We consider the single-qubit state
\begin{equation}
    \rho_{\theta} = \frac{1}{2} \left( I_2+\theta \sigma_{x}\right),
\end{equation}
where $\sigma_x$ is the Pauli matrix defined in Eq. \eqref{eq:Pauli}, and $I_2$ is the $2 \times 2$ identity matrix. The unknown parameter is $\theta \in [-1,1]$. We assume the uniform prior PDF
\begin{equation}
    \pi(\theta)=\begin{cases} \frac{1}{2}& \theta \in [-1,1], \\ 0 &\text{otherwise}. \end{cases}
\end{equation}
For a single copy ($N=1$), the operators $\Gamma^{(1)}_0$ and $\Gamma^{(1)}_1$ are given by
\begin{equation}
 \Gamma^{(1)}_0= \int^{1}_{-1} d\theta\, \rho_{\theta} \pi(\theta)=\frac{1}{2} I_2, \quad
 \Gamma^{(1)}_1= \int^{1}_{-1} d\theta\, \theta \rho_{\theta} \pi(\theta)=\frac{1}{6}\sigma_x, 
\end{equation}
Substituting these expressions into Eq.~\eqref{Solution} yields
\begin{equation}
    \Lambda_{\text{opt}}^{(1)}=\frac{1}{3}\sigma_x.
\end{equation}
Hence, the eigenvalues are $\pm \frac{1}{3}$, with the corresponding optimal PVM given by the projectors
\begin{equation}
    P^{(1)}_{\pm}=\frac{1}{2}\left(I_2 \pm \sigma_x\right).
\end{equation}
Proceeding analogously, for $N=2$ we obtain
\begin{equation}
    \Lambda_{\text{opt}}^{(2)}=\frac{1}{4}\left( I_2 \otimes \sigma_x + \sigma_x \otimes I_2 \right).
\end{equation}
The spectral decomposition of $\Lambda_{\text{opt}}^{(2)}$ yields the eigenvalues $\pm\frac{1}{2}$ and $0$, where the zero eigenvalue is twofold degenerate. The corresponding spectral projectors are
\begin{equation}
P^{(2)}_{\pm\frac{1}{2}}
=P^{(1)}_{\pm}\otimes P^{(1)}_{\pm},
\qquad
P^{(2)}_{0}
=P^{(1)}_{+}\otimes P^{(1)}_{-}
+P^{(1)}_{-}\otimes P^{(1)}_{+}.
\end{equation}
Although the eigenvectors spanning the zero-eigenspace are not uniquely defined due to the degeneracy, the projector $P^{(2)}_{0}$ onto the entire zero-eigenspace is unique.
Before discussing the general $N$-copy case, we consider the three-copy setting ($N=3$). In this case,
\begin{equation}
    \Lambda_{\text{opt}}^{(3)}=\frac{1}{5}\left( I_2 \otimes I_2\otimes \sigma_x + I_2 \otimes \sigma_x \otimes I_2+\sigma_x \otimes I_2 \otimes I_2 \right).
\end{equation}
The spectral decomposition of $ \Lambda_{\text{opt}}^{(3)}$ yields the eigenvalues $\frac{3}{5}$, $\frac{1}{5}$, $-\frac{1}{5}$, $-\frac{3}{5}$ with degeneracies $1$, $3$, $3$, and $1$, respectively. The corresponding spectral projectors are
\begin{eqnarray}
  P^{(3)}_{\pm \frac{3}{5}}=P^{(1)}_{\pm} \otimes P^{(1)}_{\pm} \otimes P^{(1)}_{\pm}, \\
  P^{(3)}_{\frac{1}{5}}=P^{(1)}_{+} \otimes P^{(1)}_{+}\otimes P^{(1)}_{-}+ P^{(1)}_{+} \otimes P^{(1)}_{-}\otimes P^{(1)}_{+}+P^{(1)}_{-} \otimes P^{(1)}_{+}\otimes P^{(1)}_{+}, \nonumber \\
  P^{(3)}_{-\frac{1}{5}}=P^{(1)}_{+} \otimes P^{(1)}_{-}\otimes P^{(1)}_{-}+ P^{(1)}_{-} \otimes P^{(1)}_{+}\otimes P^{(1)}_{-}+P^{(1)}_{-} \otimes P^{(1)}_{-}\otimes P^{(1)}_{+}. \nonumber
\end{eqnarray}
For the threefold-degenerate eigenvalues $\frac{1}{5}$ and $-\frac{1}{5}$, the individual eigenvectors within the corresponding eigenspaces are not uniquely defined. However, the spectral projectors onto the corresponding eigenspaces are unique and are given by the expressions above. The $N$-copy state is given by
\begin{equation}
    \rho_{\theta}^{\otimes N} = \frac{1}{2^N} \left( I_2+\theta \sigma_{x}\right)^{\otimes N}.
\end{equation}
Expanding the tensor product according to the number of factors containing $\sigma_x$, we obtain
\begin{equation}
    \rho_{\theta}^{\otimes N} =\frac{1}{2^{N}} \sum^N_{k=0} \theta^k A_k,
\end{equation}
where $A_k$ denotes the sum of all tensor products containing exactly $k$ factors of $\sigma_x$ and $N-k$ factors of $I_2$. For example,
\begin{equation}
    A_1= \sigma_{x} \otimes I_2 \otimes \dots \otimes I_2 +  I_2 \otimes \sigma_{x} \otimes \dots \otimes I_2 + \dots + I_2 \otimes I_2 \otimes \dots \otimes \sigma_{x}.
    \label{eq:Ifor}
\end{equation}
Using the uniform prior $\pi(\theta)$, we have
\begin{equation}
 \int^{1}_{-1} d\theta\, \theta^k \pi(\theta) =\frac{1+(-1)^k}{2(k+1)}.  
\end{equation}
Therefore, the operators entering the Lyapunov equation are
\begin{equation}
\label{eq:gamma01}
    \Gamma^{(N)}_0= \frac{1}{2^{N}} \sum^N_{k=0} \frac{1+(-1)^k}{2(k+1)} A_k \quad \text{and} \quad
    \Gamma^{(N)}_1= \frac{1}{2^{N}} \sum^N_{k=0} \frac{1+(-1)^{k+1}}{2(k+2)} A_k.
\end{equation}
The products $A_kA_1$ and $A_1A_k$ can be understood by considering the number of factors of $\sigma_x$ in each tensor product. Each term in $A_k$ contains exactly $k$ factors of $\sigma_x$ and $N-k$ factors of $I_2$. When such a term is multiplied by a term in $A_1$, there are two possibilities. If the $\sigma_x$ from $A_1$ acts on a position already containing $\sigma_x$, then $\sigma_x^2=I_2$, and the resulting term contains $k-1$ factors of $\sigma_x$. If it acts on a position containing $I_2$, the resulting term contains $k+1$ factors of $\sigma_x$. To determine the corresponding coefficients, consider a fixed term in $A_{k+1}$. It can be obtained in $k+1$ different ways when terms of $A_k$ and $A_1$ are multiplied. Thus, the contribution from the terms containing $k+1$ factors of $\sigma_x$ is $(k+1)A_{k+1}$. Similarly, a fixed term in $A_{k-1}$ can be obtained in $N-k+1$ different ways, since the additional $\sigma_x$ in the corresponding term of $A_k$ can occupy any of the $N-k+1$ positions that contain $I_2$ in the final term. Hence,
\begin{eqnarray}
&&A_0 A_1 = A_1 A_0 =A_1, \\
&&A_k A_1 = A_1 A_k=(k+1) A_{k+1} + (N-k+1) A_{k-1},
\quad k=1,\dots,N-1, \nonumber \\
&&A_N A_1 = A_1 A_N=A_{N-1}. \nonumber
\end{eqnarray}
These relations imply that $A_1$ commutes with all operators $A_k$. Since $\Gamma_0^{(N)}$ and $\Gamma_1^{(N)}$ are linear combinations of the operators $A_k$, it follows that
\begin{equation}
\label{eq:commutations}
\left[A_1,\Gamma_0^{(N)}\right]=
\left[A_1,\Gamma_1^{(N)}\right]=0.    
\end{equation}
Furthermore, we have
\begin{equation}
    A_1 \Gamma^{(N)}_0= \frac{1}{2^{N}} \sum^N_{k=0} \frac{1+(-1)^k}{2(k+1)} \left[(k+1) A_{k+1} + (N-k+1) A_{k-1}\right],
\end{equation}
where we use the convention $A_{-1}=A_{N+1}=0$. Separating the two terms gives
\begin{equation}
 A_1 \Gamma^{(N)}_0= \frac{1}{2^{N}} \sum^N_{k=0} \left[(k+1)\frac{1+(-1)^k}{2(k+1)} A_{k+1} + (N-k+1) \frac{1+(-1)^k}{2(k+1)} A_{k-1}\right]. 
\end{equation}
In the first sum we make the substitution $j=k+1$, while in the second sum we use $j=k-1$. Then
\begin{equation}
 A_1 \Gamma^{(N)}_0= \frac{1}{2^{N}} \sum^N_{j=0} \left[j\frac{1+(-1)^{j-1}}{2j} + (N-j) \frac{1+(-1)^{j+1}}{2(j+2)} \right]A_j.   
\end{equation}
Because
\begin{equation}
  1+(-1)^{j-1}= 1+(-1)^{j+1}, 
\end{equation}
the coefficient of $A_j$ simplifies to
\begin{eqnarray}
 j\frac{1+(-1)^{j-1}}{2j} + (N-j) \frac{1+(-1)^{j+1}}{2(j+2)}&=& \frac{1+(-1)^{j+1}}{2} \left[1+\frac{N-j}{j+2}\right]  \nonumber \\
 &=& (N+2)\frac{1+(-1)^{j+1}}{2(j+2)}.
\end{eqnarray}
Therefore,
\begin{equation}
   A_1 \Gamma^{(N)}_0=\frac{N+2}{2^{N}} \sum^N_{j=0} \frac{1+(-1)^{j+1}}{2(j+2)}A_j. 
\end{equation}
Comparing this expression with the definition of $\Gamma^{(N)}_1$ in \eqref{eq:gamma01}, we obtain
\begin{equation}
    A_1 \Gamma^{(N)}_0=(N+2) \Gamma^{(N)}_1,
\end{equation}
which, together with the commutation relations in \eqref{eq:commutations}, yields
\begin{equation}
    \left[\Gamma^{(N)}_0, \Gamma^{(N)}_1\right]=0.
\end{equation}
This also implies that Eq. \eqref{Solution} simplifies to
\begin{eqnarray}
    \Lambda_{\text{opt}}^{(N)} &=& 2 \int_{0}^{\infty} e^{-\Gamma_{0}^{(N)} x} \Gamma_{1}^{(N)} e^{-\Gamma_{0}^{(N)} x} dx=\Gamma_{1}^{(N)} \left(\Gamma_{0}^{(N)}\right)^{-1} \nonumber \\
    &=&\frac{1}{N+2}A_1\Gamma_{0}^{(N)} \left(\Gamma_{0}^{(N)}\right)^{-1}=\frac{1}{N+2}A_1,
    \label{eq:IIfor}
\end{eqnarray}
where we have used the fact that $\Gamma_{0}^{(N)}$ is a positive definite operator. 

To obtain the spectral decomposition of $ \Lambda_{\text{opt}}^{(N)}$, we rewrite the terms in $A_1$ according to
\begin{equation}
    \sigma_x=P^{(1)}_{+}-P^{(1)}_{-} \quad \text{and} \quad I_2=P^{(1)}_{+}+P^{(1)}_{-}.
\end{equation}
Thus, every tensor-product projector
\begin{equation}
    P^{(1)}_{s_1} \otimes P^{(1)}_{s_2} \otimes \dots \otimes P^{(1)}_{s_N}, \quad s_i \in \{+,-\}
\end{equation}
is an eigenprojector of $\Lambda_{\text{opt}}^{(N)}$.

Furthermore, the eigenvalues of $\Lambda_{\text{opt}}^{(N)}$ can be deduced in the following manner. We consider $\Lambda_{\text{opt}}^{(N)}$ being acted on a tensor-product projector containing $k$ factors $P^{(1)}_{+}$ and $N-k$ factors $P^{(1)}_{-}$. Recall that there can only be one $\sigma_{x}=P_{+}^{(1)}-P_{-}^{(1)}$ in every tensor-product term in the above expression of $\Lambda_{\text{opt}}^{(N)}$, see Eqs. \eqref{eq:Ifor} and \eqref{eq:IIfor}.
If the tensor-product projector has a $P_+^{(1)}$ ($P_-^{(1)}$) at the position corresponding to the $\sigma_x$ in a given term of $\Lambda_{\text{opt}}^{(N)}$, then that term contributes a factor of $+1$ ($-1$). Since there are $N$ terms in $\Lambda_{\text{opt}}^{(N)}$, the contribution of a given tensor-product projector is obtained by considering all $N$ possible positions of $\sigma_x$. As an example,  
$P^{(1)}_{-} \otimes P^{(1)}_{-} \otimes \dots \otimes P^{(1)}_{-}$ leads to a factor of $-1$ for each of the $N$ terms. Therefore, the corresponding eigenvalues are
\begin{equation}
   \frac{k-(N-k)}{N+2}=\frac{2k-N}{N+2}, \quad k=0,1, \dots, N.
\end{equation}
For a fixed $k$, there are $\binom{N}{k}$ different tensor-product projectors containing exactly $k$ factors $P^{(1)}_{+}$. Hence, the eigenvalue $\frac{2k-N}{N+2}$ has degeneracy $\binom{N}{k}$. The corresponding spectral projector, projecting on a $\binom{N}{k}$-dimensional subspace, is obtained by summing all tensor-product projectors containing exactly $k$ factors $P^{(1)}_{+}$ and $N-k$ factors $P^{(1)}_{-}$. Thus, the spectral projectors are given by
\begin{eqnarray}
  P^{(N)}_{-\frac{N}{N+2}}&=P^{(1)}_{-} \otimes P^{(1)}_{-} \otimes \dots \otimes P^{(1)}_{-}, \\
  P^{(N)}_{-\frac{N-2}{N+2}}&=P^{(1)}_{+} \otimes P^{(1)}_{-} \otimes \dots \otimes P^{(1)}_{-} +  P^{(1)}_{-} \otimes P^{(1)}_{+} \otimes \dots \otimes P^{(1)}_{-} \nonumber \\&+ \dots + P^{(1)}_{-} \otimes P^{(1)}_{-} \otimes \dots \otimes P^{(1)}_{+}, \nonumber \\
\dots \dots& \dots \dots \dots \dots \nonumber \\
  P^{(N)}_{\frac{N}{N+2}}&=P^{(1)}_{+} \otimes P^{(1)}_{+} \otimes \dots \otimes P^{(1)}_{+}. \nonumber
\end{eqnarray}
We therefore conclude that, in general, the projectors of the optimal PVM are not guaranteed to be in form $P_1\otimes P_2\otimes \dots \otimes P_N$. Instead, for degenerate eigenvalues, the projection onto the corresponding eigenspace is obtained as the sum of all tensor-product projectors containing a fixed number of factors $P^{(1)}_{-}$ and $P^{(1)}_{+}$. The resulting spectral projectors are uniquely defined by the corresponding eigenspaces, although the choice of basis within a degenerate eigenspace is not unique. Only the projectors associated with the largest eigenvalue and the smallest eigenvalue factorize as a single tensor product.

\end{document}